\documentclass[letterpaper,twocolumn,nofootinbib,showkeys]{revtex4}
\usepackage{mysty}
\usepackage{graphicx}
\usepackage{hyperref}
\usepackage{multirow}
\usepackage[sort&compress]{natbib}
\bibpunct{(}{)}{;}{a}{}{,}

\begin{document}

\title{Large Tandem, Higher Order Repeats and Regularly Dispersed
Repeat Units Contribute Substantially to Divergence Between Human
and Chimpanzee Y Chromosomes}

\author{V. Paar$^{\dag}$}\email{paar@hazu.hr}
\author{M. Glun\v{c}i\'c\footnote{V. Paar and M. Glun\v{c}i\'c contributed equally to this work.}}
\author{I. Basar}
\author{M. Cvitkovi\'c}
\affiliation{Faculty of Science, University of Zagreb, 10000
Zagreb, Croatia}
\author{M. Rosandi\'c}
\affiliation{Department of Internal Medicine, University Hospital
Rebro, University of Zagreb, 10000 Zagreb, Croatia}
\author{P. Paar}
\affiliation{Faculty of Electrical Engineering and Computing,
10000 Zagreb, Croatia}

\date{\today}

\begin{abstract}
Comparison of human and chimpanzee genomes has received much
attention, because of paramount role for understanding
evolutionary step distinguishing us from our closest living
relative. In order to contribute to insight into Y chromosome
evolutionary history, we study and compare tandems, higher order
repeats (HORs), and regularly dispersed repeats in human and
chimpanzee Y chromosome contigs, using robust Global Repeat Map
algorithm. We f\mbox{}ind a new type of long--range acceleration,
human--accelerated HOR regions. In peripheral domains of 35mer
human alphoid HORs, we f\mbox{}ind riddled features with ten
additional repeat monomers. In chimpanzee, we identify 30mer
alphoid HOR. We construct alphoid HOR schemes showing
signif\mbox{}icant human--chimpanzee dif\mbox{}ference, revealing
rapid evolution after human--chimpanzee separation. We identify
and analyze over 20 large repeat units, most of them reported here
for the f\mbox{}irst time as: chimpanzee and human $\sim$1.6 kb
3mer secondary repeat unit (SRU) and $\sim$23.5 kb tertiary repeat
unit ($\sim$0.55 kb primary repeat unit, PRU); human 10848, 15775,
20309, 60910, and 72140 bp PRUs; human 3mer SRU ($\sim$2.4 kb
PRU); 715mer and 1123mer SRUs (5mer PRU); chimpanzee 5096, 10762,
10853, 60523 bp PRUs; and chimpanzee 64624 bp SRU (10853 bp PRU).
We show that substantial human--chimpanzee dif\mbox{}ferences are
concentrated in large repeat structures, at the level of as much
as $\sim$70\% divergence, sizably exceeding previous numerical
estimates for some selected noncoding sequences. Smeared over the
whole sequenced assembly (25 Mb) this gives $\sim$14\%
human–-chimpanzee divergence. This is signif\mbox{}icantly higher
estimate of divergence between human and chimpanzee than previous
estimates.
\end{abstract}

  \maketitle

\section{Introduction}

\subsection{Atypical Structure of Human Y Chromosome}

One of challenging problems in genomics is related to the
evolutionary development of Y chromosome. The Y chromosome has a
unique role in human population genetics with properties that
distinguish it from all other chromosomes
\citep{mitchell,jobling,skaletsky}. Prevailing theory is that X
and Y chromosomes evolved from a pair of autosomes
\citep{muller,ohno,graves,lahn,marshall}. Lack of recombination
between nonrecombining parts of X and Y chromosomes was thought to
be responsible for decay of the Y--linked genes, the pace of which
slows over time, eventually leading to a paucity of genes.
Identif\mbox{}ication of distinct palindromes harboring several
distinct gene families unique to the long arm of Y chromosome,
frequent gene conversion, and multiplication have raised some
doubt about progressive decay of the Y chromosome
\citep{kuroda,skaletsky,rozen,ali,knijff}. It was shown that the Y
chromosome has acquired a large number of testis specif\mbox{}ic
genes during the course of evolution, including those essential
for spermatogenesis \citep{saxena96,silber,skaletsky}.

Considerations of atypical structure of human Y chromosome were
largely focused on the gene--related content. On the other hand,
however, the human Y chromosome is replete with many pronounced
repetitive sequences, and multicopy gene arrays are embedded in
palindromes
\citep{tyler85,wolfe,tyler87,oakey,cooper1,skaletsky,rozen,perry07,kircsh}.

\subsection{Alphoid Higher Order Repeats}

Alphoid arrays in centromeres of human and other mammal chromosomes consist
of tandem repeats of AT--rich alpha satellites
\citep{maio,manuelidis,mitchell,tyler85,willard85,waye,tyler87,romanova,warburton96a,warburton96b,choo,alexandrov,rudd06}.
Stretches of alpha satellites lacking any higher--order periodicity mutually
diverge by$\sim$20–-35\% and are referred to as monomeric
\citep{warburton96a}.

Higher order repeats (HORs) are def\mbox{}ined as higher order periodicity
pattern superimposed on the approximately periodic tandem of alpha monomers:
if an array of $n$ monomers denoted by $1, 2,\dots ,n$ is followed by the
next array of monomers denoted by $n + 1, n + 2, \dots ,2n$, where the
monomer 1 is almost identical (more than 95\%) to the monomer $n + 1$, the
monomer 2 to the monomer $n + 2$, and the monomer $n$ to the monomer $2n$,
these arrays belong to the $n$mer HOR \citep{warburton96a}. The HOR copies
from the same locus diverge from each other by $< 5$\%, while the alpha
satellite copies within any HOR copy diverge from each other by $\sim
20-35$\% \citep{warburton96a}.

Alphoid HORs are chromosome--specif\mbox{}ic
\citep{willard85,jorgensen86,willard87,haaf92,warburton96a,choo}.
A type of polymorphism found in alphoid arrays involves HOR units
that dif\mbox{}fer by an integral number of monomers (monomer
insertion or deletion), but nonetheless closely related in
sequence \citep{haaf92,warburton96a}.

Investigations using restriction endonuclease digestion have
revealed a major block of alphoid DNA in the centromeric region of
human Y chromosome
\citep{mitchell,tyler85,wolfe,tyler87,cooper1,cooper2}. The size
of this alphoid block was found to be polymorphic, widely varying
between dif\mbox{}ferent individuals \citep{tyler87,oakey}.
Initially, a 5.7 kb HOR unit was reported as a major variant of
secondary periodicity and 6.0 kb HOR unit as a minor variant.
These HOR units were associated with 34mer and 36mer, respectively
\citep{tyler87}. In a more recent study, a 5941 bp secondary
periodicity (35 alphoid repeat units) was reported
\citep{skaletsky}.

The alpha satellite DNA can be considered as a paradigm for
processes of concerted evolution in tandemly repeated DNA families
\citep{willard87,willard91,warburton96a}.

\subsection{Bioinformatics Studies of Alphoid HORs}

During the last decade sequence contigs spanning the junction at
the edges of the centromere DNA array are available for
bioinformatics analyses
\citep{rudd03,skaletsky,rosandic03a,rosandic03b,rosandic06,rudd04,paar05,paar07,ross,nusbaum}.
However, major gaps still remain at the centromeric region of
chromosomes \citep{schueler,henikoff,rudd04}. Mostly, only
peripheral HOR copies are accessible, at the edges of centromeric
region. Previously, \citet{rudd04} analyzed the Build 34 assembly,
using a combination of BLAST \citep{altschul} and DOTTER
\citep{sonnhammer}, and reported the presence of HORs. Recently,
using Tandem Repeat F\mbox{}inder (TRF) \citep{benson} and other
standard bioinformatics tools, \citet{gelfand} and
\citet{warburton08} studied human HORs in more details.

In a dif\mbox{}ferent approach, we have shown that the Key String
Algorithm (KSA) and an extension Global Repeat Map (GRM) are
ef\mbox{}fective in identif\mbox{}ication and analysis of
intrinsic structure of HORs
\citep{rosandic03a,rosandic03b,rosandic06,paar05,paar07}. Applying
KSA and GRM to the NCBI human genome assembly, detailed structure
of known and some new human alphoid HORs was determined.

\subsection{Comparison of Human and Chimpanzee Genome
Sequences}

To understand the genetic basis of unique human features, the
human and chimpanzee genomes have been compared in a number of
studies
\citep{sibley,laursen,king,haaf97,haaf98,chen,pennacchio,fujiyama,
olson,boffelli,webster,watanabe,cheng,khaitovich,mikkelsen,newman,
bailey,patterson,varki05,kuroki,ebersberger07,kehrer,perry08,varki08,liu}.
Large variation in sequence divergence was often seen among
genomic regions. For example, the last intron of the ZFY gene
showed only 0.69\% divergence between human and chimpanzee
\citep{dorit}, whereas for the OR1D3P pseudogene a divergence of
3.04\% was found \citep{glusman}. Thus, to have reliable estimates
of the average divergences between hominoid genomes, it was
concluded that sequence data from many genomic regions are needed
\citep{chen}. Estimates of divergence due to nucleotide
substitutions were about 1.24\% between selected intergenic
nonrepetitive DNA segments in humans and chimpanzees,
substantially lower than previous ones, of about 3\%, which
included repetitive sequences
\citep{chen,fujiyama,ebersberger02,mikkelsen}. A greater sequence
divergence (1.78\%) was obtained between reported f\mbox{}inished
sequence of the chimpanzee Y chromosome (PTRY) and the human Y
chromosome \citep{kuroki}. Comparing the DNA sequences of unique,
Y--linked genes in chimpanzee and human, evidence was found that
in the human lineage all such genes were conserved, and in the
chimpanzee lineage, by contrast, several genes have sustained
inactivating mutations \citep{hughes05}.

On the other hand, the overall sequence divergence by taking
regions of indels into account was estimated to be approximately
5\% \citep{britten1,britten2,cheng,gibbs}. In some short stretches
of human and chimpanzee genomes, so called human--accelerated
regions, signif\mbox{}icant increase of substitution divergence
was found
\citep{pollard06a,pollard06b,popesco,prabhakar,pollard09}. On the
other hand, based on phylogenetic analysis of large number of DNA
sequence alignments from human and chimpanzee it was found that
for a sizeable fraction of our genome we share no immediate
genetic ancestry with chimpanzee \citep{ebersberger07}.

Experimental evidence suggests that a progenitor of
suprachromosomal alphoid family 3 was established and dispersed to
chimpanzee chromosomes homologous to human chromosomes 1, 11, 17
and X prior to the human--chimpanzee split
\citep{durfy90,baldini,willard91,warburton96b}. Notably, the
alphoid HOR organization in the X chromosome has been conserved
\citep{durfy90}; only the localization of the suprachromosomal
family (SF) 3 alpha satellite is substantially conserved. It was
concluded that the lack of sequence or HOR conservation among
human and chimpanzee indicates that most alpha satellite sequences
do not evolve orthologously.

In a recent publication, \citet{hughes10} have shown by sequence
comparison of human and chimpanzee MSY that humans and chimpanzees
dif\mbox{}fer radically in sequence structure and gene content. It
was concluded that, since the separation of human and chimpanzee
lineages, sequence gain and loss have been far more concentrated
in the MSY than in the balance of the genome, indicating
accelerated structural remodeling of the MSY in the chimpanzee and
human lineages during the past six million years.

The previously reported 35mer alphoid HOR in human Y chromosome
\citep{tyler87,warburton96a,skaletsky} involves the largest
alphoid HOR unit found in human genome and it is of particular
interest to look for divergence between alphoid HOR in human and
chimpanzee Y chromosome. Alphoid HOR in chimpanzee Y chromosome
was not yet reported.

Having in mind possibly important information regarding the
evolutionary role of human and chimpanzee Y chromosomes and
availability of their genomic sequences
\citep{skaletsky,mikkelsen} and a demanding task of studying
bioinformatically such long HOR units, we perform here an
extensive study applying novel robust bioinformatics tools GRM. We
investigate the major alphoid HOR from Build 37.1 assembly of
human Y chromosome and determine detailed monomer scheme and
consensus sequence, f\mbox{}inding a riddling pattern not reported
previously. In the chimpanzee Y chromosome, for the f\mbox{}irst
time, we identify and analyze alphoid HOR. We f\mbox{}ind that the
human and chimpanzee HORs are sizeably dif\mbox{}ferent, both in
size and composition of HOR units and in the constituting monomer
structure.

Furthermore, we identify and investigate in human and chimpanzee Y
chromosomes more than 20 other tandems, HORs and regularly
dispersed repeats based on large repeat units, showing sizeable
human--chimpanzee divergence. Most of these repeats are reported
here for the f\mbox{}irst time.

\section{Materials and Methods}

\subsection{Key String Algorithm}

In spite of powerful standard bioinformatics tools, there are
still dif\mbox{}f\mbox{}iculties to identify and analyze large
repeat units. For example, the detection limit of TRF is 2 kb
\citep{gelfand,warburton08}. Here, we use a new approach useful in
particular for very long and/or complex repeats.

The KSA framework is based on the use of a freely chosen short
sequence of nucleotides, called a key string, which cuts a given
genomic sequence at each location of the key string within genomic
sequence. Going along genomic sequence, the lengths of ensuing KSA
fragments form KSA length array. Such array could be compared to
an array of lengths of restriction fragments resulting from a
hypothetical complete digestion, cutting genomic sequence at
recognition sites corresponding to KSA key string. Any periodicity
appearing in the KSAlength array enables identif\mbox{}ication and
location of repeat in a given genomic sequence. Analysis of repeat
sequences at position of any periodicity in the KSA length array
gives consensus repeat unit and divergence of each repeat copy
with respect to consensus. Any presence of higher order
periodicity in the KSA length array reveals the presence of HOR at
that location and enables determination of consensus HOR repeat
unit and divergence of each HOR copy with respect to consensus.

Similarly, with a proper choice of key string, the KSA fragments a
given tandem repeat into monomers, as for example cutting Alu
sequence at two identical positions providing
identif\mbox{}ication of Alu sequences, cuts a palindrome
providing identif\mbox{}ication of large palindrome sequences and
their substructure, and so on. KSA provides a straightforward
ordering of KSA fragments, regardless of their size (from small
fragments of a few bp to as large as tens of kilobasepairs). KSA
provides high degree of robustness and requires only a modest
scope of computations using PC. Due to its robustness, KSA is
ef\mbox{}fective even in cases of signif\mbox{}icant deletions,
insertions, and substitutions, providing detailed HOR annotation
and structure, consensus sequence, and exact consensus length in a
given genomic sequence even if it is highly distorted, intertwined
and riddled (segmentally fuzzy repeats). Using a HOR consensus
sequence, in the next step KSA computes f\mbox{}iner
characteristics, as for example the SF classif\mbox{}ication and
CENP--B box/pJ$\alpha$ distributions.

\subsection{Global Repeat Map}

The GRM program is an extension of KSA framework. GRM of a given
genomic sequence is executed in f\mbox{}ive steps.
\begin{list}{\labelitemi}{\leftmargin=2.5em}
    \item[Step 1] \emph{GRM--Total module} Computes the frequency versus fragment
length distribution for a given genomic sequence by superposing
results of consecutive KSA segmentations computed for an ensemble
of all 8 bp key strings ($4^8 = 65536$ key strings). In GRM
diagram, each pronounced peak corresponds to one or more repeats
at that length, tandem or dispersed. GRM computation is fast and
can be easily executed for human chromosome using PC.
    \item[Step 2] \emph{GRM--Dom module} Determines dominant key string
corresponding to fragment length for each peak in the GRM diagram
from the step 1. A particular 8 bp key string (or a group of 8 bp
key strings) that gives the largest frequency for a fragment
length under consideration is referred to as dominant key string.
    \item[Step 3] \emph{GRM--Seg module} Performs segmentation of a given genomic
sequence into KSA fragments using dominant key string from the
step 2. Any periodic segment within the KSA length array reveals
the location of repeat and provides genomic sequences of the
corresponding repeat copies.
    \item[Step 4] \emph{GRM--Cons module} Aligning all sequences of repeat copies
from step 3 constructs the consensus sequence.
    \item[Step 5] \emph{NW module} Computes divergence between each repeat copy from
step 3 and consensus sequence from step 4 using Needleman--Wunsch
algorithm \citep{needleman}.
\end{list}

Regarding the 8 bp choice of key string size: using an ensemble of
$r$--bp key strings the average length of KSA fragments is $\sim
4^r$. With increasing length of key strings the overall frequency
of large fragment lengths increases. We tested that the 8 bp key
string ensemble is suitable for identif\mbox{}ication of repeat
units in a wide range of lengths, from $\sim$10 bp to as much as
$\sim$100 kb. However, from GRM construction it follows that fully
reliable results are obtained for key string lengths not exceeding
the repeat length under study.

In summary, the characteristics of GRM are:
\begin{list}{\labelitemi}{\leftmargin=1em}
    \item[--] robustness of the method with respect to deviations from perfect
repeats, i.e., substitutions, insertions, and deletions;
    \item[--] use of ensemble of all 8 bp key strings as a starting point of
algorithm, thus avoiding the need to choose a particular key
string for any repeat structure;
    \item[--] straightforward identif\mbox{}ication of repeats (tandem and
dispersed), applicable to very large repeat units, as large as
tens of kilobasepairs;
    \item[--] easy identif\mbox{}ication of HORs and determination of consensus
lengths and consensus sequences.
\end{list}

\section{Results and Discussion}

Using GRM algorithm we have identif\mbox{}ied and analyzed tandem
repeats, HORs and regularly dispersed repeats with large repeat
units in human and chimpanzee Y chromosomes (Build 37.1 and Build
2.1 assemblies, respectively). Summary of all large repeat units
identif\mbox{}ied and analyzed in this article and the
human--chimpanzee comparison are given in Tables \ref{t1},
\ref{t2}, and \ref{t3}.
\begin{table*}[t]
\centering \caption{\label{t1}\textsf{\small{Tandem repeats, HORs
and dispersed repeats with large repeat units in contigs of human
Y chromosome.}}}
\begin{tabular}{lp{0.51cm}lp{0.51cm}lp{0.51cm}lp{0.51cm}lp{0.51cm}lp{0.51cm}l}
\hline \multirow{2}{*}{Repeat unit (bp)} & &
\multirow{2}{*}{Structure} & & \multirow{2}{*}{Character} & &
\multirow{2}{*}{Contig} & &
Chr Y start & & Chr Y end & & \multirow{2}{*}{Length} \\
                 & &           & &           & &        & & position                     & & position                   & &        \\
\hline
$\sim$171                       & & PRU\footnotemark[1]  & & Alpha satellite       & & NT\_011878.9   & & 10083775  & & 13131913  & & 3048138   \\
35mer(45mer\footnotemark[3])    & & SRU\footnotemark[1]  & & Alphoid HOR           & & NT\_087001.1   & &           & &           & &           \\
125                             & & PRU                  & & Tandem                & & NT\_011875.12  & & 22216726  & & 22513032  & & 296306    \\
$\sim$545                       & & PRU\footnotemark[1]  & & Tandem                & & see Table \ref{t5} & &       & &           & & 12577     \\
$\sim$1641\footnotemark[3]      & & SRU\footnotemark[1]  & & Regularly dispersed   & &                & &           & &           & &           \\
$\sim$23541\footnotemark[3]     & & TRU\footnotemark[2]  & & Third order tandem    & & NT\_011903.12  & & 24023693  & & 24070760  & & 47067     \\
                                & &                      & &                       & &                & & 24312159  & & 24333896  & & 21737     \\
                                & &                      & &                       & &                & & 24544818  & & 24566560  & & 21742     \\
                                & &                      & &                       & & NT\_011875.12  & & 23654663  & & 23713744  & & 59081     \\
$\sim$2385                      & & PRU\footnotemark[1]  & & Tandem                & & NT\_011903.12  & & 25298078  & & 25312458  & & 14380     \\
$\sim$4757\footnotemark[3]      & & SRU\footnotemark[1]  & & 2mer HOR              & &                & & 25376692  & & 25424719  & & 48027     \\
$\sim$7155\footnotemark[3]      & & SRU\footnotemark[1]  & & 3mer HOR              & &                & & 26929417  & & 26948531  & & 19114     \\
                                & &                      & &                       & &                & & 27001927  & & 27038009  & & 36082     \\
5                               & & PRU\footnotemark[1]  & & Tandem                & & NT\_025975.2   & & 58819393  & & 58917657  & & 98264     \\
$\sim$3579                      & & SRU\footnotemark[1]  & & 715mer HOR            & &                & &           & &           & &           \\
5                               & & PRU\footnotemark[1]  & & Tandem                & & NT\_113819.1   & & 13690637  & & 13747836  & & 57199     \\
$\sim$5607\footnotemark[3]      & & SRU\footnotemark[1]  & & 1123mer HOR           & &                & &           & &           & &           \\
$\sim$5096\footnotemark[3]      & & PRU\footnotemark[2]  & & Dispersed             & & NT\_011875.12  & & 20121395  & & 20126501  & & 5106      \\
                                & &                      & &                       & &                & & 20003268  & & 20008374  & & 5106      \\
                                & &                      & & Dispersed             & & NT\_011903.12  & & 26206614  & & 26211701  & & 5087      \\
                                & &                      & &                       & &                & & 27750731  & & 27755818  & & 5087      \\
$\sim$10848                     & & PRU\footnotemark[2]  & & Tandem                & & NT\_011903.12  & & 25312733  & & 25341062  & & 28329     \\
                                & &                      & &                       & &                & & 26984151  & & 27001645  & & 17494     \\
$\sim$15766\footnotemark[3]     & & PRU\footnotemark[2]  & & Dispersed             & & NT\_011875.12  & & 23167813  & & 23183579  & & 15766     \\
                                & &                      & &                       & &                & & 23209651  & & 23225434  & & 15783     \\
$\sim$15775\footnotemark[3]     & & PRU\footnotemark[2]  & & Tandem                & & NT\_011896.9   & & 6543373   & & 6574923   & & 31550     \\
                                & & PRU\footnotemark[2]  & & Dispersed             & & NT\_011651.17  & & 14540408  & & 14556183  & & 15775     \\
$\sim$20309                     & & PRU\footnotemark[2]  & & Tandem                & & NT\_011878.9   & & 9293306   & & 9374535   & & 81229     \\
                                & & PRU\footnotemark[2]  & & Tandem                & & NT\_086998.1   & & 9170808   & & 9241328   & & 70520     \\
$\sim$60910\footnotemark[3]     & & PRU\footnotemark[2]  & & Dispersed             & & NT\_011875.12  & & 19697222  & & 19759044  & & 60917     \\
                                & &                      & &                       & &                & & 20420735  & & 20482553  & & 60909     \\
$\sim$72140\footnotemark[3]     & & PRU\footnotemark[2]  & & Dispersed             & & NT\_011875.12  & & 19829682  & & 19900381  & & 70699     \\
                                & &                      & &                       & &                & & 20279397  & & 20350098  & & 70701     \\
\hline
\end{tabular}
\footnotetext[1]{Described in text} \footnotetext[2]{Described in
Supplementary text} \footnotetext[3]{For the f\mbox{}irst time
reported in this work}

\begin{minipage}[t]{\textwidth}
\raggedright \footnotesize{\emph{PRU} primary repeat unit,
\emph{SRU} secondary repeat unit, \emph{TRU} tertiary repeat unit,
\emph{dispersed} dispersed at random spacings, \emph{regularly
dispersed} dispersed at regular spacings}
\end{minipage}
\end{table*}

\begin{table*}[t]
\centering \caption{\label{t2}\textsf{\small{Tandem repeats, HORs
and dispersed repeats with large repeat units in contigs of
chimpanzee Y chromosome.}}}
\begin{tabular}{lp{0.45cm}lp{0.45cm}lp{0.45cm}lp{0.45cm}lp{0.45cm}lp{0.45cm}l}
\hline \multirow{2}{*}{Repeat unit (bp)} & &
\multirow{2}{*}{Structure} & & \multirow{2}{*}{Character} & &
\multirow{2}{*}{Contig} & &
Chr Y start & & Chr Y end & & \multirow{2}{*}{Length} \\
& &           & &           & &        & & position                     & & position                   & &        \\
\hline
$\sim$171                   & & PRU\footnotemark[1] & & Alpha satellite     & & NW\_001252921.1 & & 7108946     & & 7151404     & & 42458   \\
30mer\footnotemark[3]       & & SRU\footnotemark[1] & & Alphoid HOR         & &                 & &             & &             & &         \\
$\sim$550\footnotemark[3]   & & PRU\footnotemark[1] & & Tandem See Table \ref{t4}  & &          & &             & &             & & 30832   \\
$\sim$1652\footnotemark[3]  & & SRU\footnotemark[1] & & Regularly dispersed & &                 & &             & &             & &         \\
$\sim$23578\footnotemark[3] & & TRU\footnotemark[1] & & Third order tandem  & & NW\_001252921.1 & & 7707476     & & 7728531     & & 21055   \\
                            & &                     & &                     & &                 & & 8130226     & & 8160370     & & 30144   \\
                            & &                     & &                     & &                 & & 8433315     & & 8464030     & & 30715   \\
                            & &                     & &                     & &                 & & 8866559     & & 8897264     & & 30705   \\
                            & &                     & &                     & &                 & & 9166900     & & 9197050     & & 30150   \\
                            & &                     & &                     & &                 & & 9598779     & & 9628923     & & 30144   \\
$\sim$2383                  & & PRU\footnotemark[1] & & Tandem              & & NW\_001252917.1 & & 3256815     & & 3278585     & & 21770   \\
                            & &                     & &                     & &                 & & 3406716     & & 3428486     & & 21770   \\
                            & &                     & & Tandem              & & NW\_001252922.1 & & 11224963    & & 11256302    & & 31339   \\
                            & &                     & &                     & &                 & & 11298117    & & 11327074    & & 28957   \\
$\sim$5096\footnotemark[3]  & & PRU\footnotemark[2] & & Tandem              & & NW\_001252916.1 & & 1956128     & & 1981606     & & 25478   \\
                            & &                     & &                     & &                 & & 2082379     & & 2092569     & & 10190   \\
                            & &                     & & Dispersed           & & NW\_001252920.1 & & 5633270     & & 5638363     & & 5093    \\
                            & &                     & & Dispersed           & & NW\_001252924.1 & & 12174453    & & 12179546    & & 5093    \\
                            & &                     & &                     & &                 & & 12280382    & & 12285475    & & 5093    \\
$\sim$10762\footnotemark[3] & & PRU\footnotemark[2] & & Tandem              & & NW\_001252919.1 & & 276373      & & 308349      & & 31976   \\
                            & &                     & & Tandem              & & NW\_001252921.1 & & 2823896     & & 2845204     & & 21308   \\
                            & &                     & & Dispersed           & & NW\_001252925.1 & & 1219588     & & 1230035     & & 10447   \\
$\sim$10853\footnotemark[3] & & PRU\footnotemark[1] & & Tandem              & & NW\_001252917.1 & & 1130756     & & 1160123     & & 29367   \\
$\sim$64624\footnotemark[3] & & SRU\footnotemark[1] & &                     & &                 & & 1174942     & & 1224747     & & 49805   \\
$\sim$60523\footnotemark[3] & & PRU\footnotemark[2] & & Tandem              & & NW\_001252918.1 & & 3827479     & & 3948523     & & 121044  \\
                            & & PRU\footnotemark[2] & & Dispersed           & & NW\_001252922.1 & & 10310933    & & 10371414    & & 60481   \\
                            & & PRU\footnotemark[2] & & Dispersed           & & NW\_001252919.1 & & 5301324     & & 5361771     & & 60447   \\
$\sim$71778\footnotemark[3] & & PRU                 & & Dispersed           & & NW\_001252925.1 & & 12394038    & & 12465796    & & 71758   \\
                            & & PRU                 & & Dispersed           & & NW\_001252915.1 & & 1775843     & & 1847647     & & 71804   \\
                            & & PRU                 & & Dispersed           & & NW\_001252917.1 & & 2201698     & & 2273485     & & 71787   \\
                            & & PRU                 & & Dispersed           & & NW\_001252919.1 & & 5440228     & & 5505887     & & 65659   \\
$\sim$72140\footnotemark[3] & & PRU                 & & Tandem              & & NW\_001252923.1 & & 11947703    & & 12091619    & & 143916  \\
\hline
\end{tabular}
\footnotetext[0]{\hspace{-\parindent}For description see Table
\ref{t1}}
\end{table*}

\begin{table}[t]
\centering \caption{\label{t3}\textsf{\small{Correspondence of
large repeat and HOR units in Y chromosome contigs of human and
chimpanzee.}}}
\begin{tabular}{ll}
\hline Human & Chimpanzee \\
\hline
125 bp PRU                                   & –-                  \\
$\sim$171 bp PRU                             & $\sim$171 bp PRU    \\
35mer/45mer SRU                              & 30mer SRU           \\
$\sim$545 bp PRU                             & $\sim$550 bp PRU    \\
$\sim$1641 bp SRU                            & $\sim$1652 bp SRU   \\
$\sim$23541 bp TRU                           & $\sim$23578 bp TRU  \\
$\sim$2385 bp PRU                            & $\sim$2383 bp PRU   \\
$\sim$4757 bp SRU                            & –-                  \\
$\sim$7155 bp SRU                            & –-                  \\
5 bp PRU                                     & 5 bp PRU            \\
$\sim$3579 bp SRU                            & –-                  \\
$\sim$5096 bp PRU (dispersed)                & $\sim$5096 bp PRU   \\
5 bp PRU                                     & 5 bp PRU            \\
5607 bp SRU                                  & –-                  \\
$\sim$10.8 kb PRU (within $\sim$20309 bp PRU)&$\sim$10762 bp PRU   \\
$\sim$10848 bp PRU                           & $\sim$10853 bp PRU  \\
–-                                           & $\sim$64624 bp SRU  \\
$\sim$15766 bp PRU (dispersed)               & Dispersed           \\
                                             & fragments           \\
$\sim$15775 bp PRU                           & -–                  \\
$\sim$60910 bp PRU (dispersed)               & $\sim$60523 bp PRU  \\
$\sim$72140 bp PRU (dispersed)               & $\sim$72140 bp PRU  \\
                                             & $\sim$71778 bp PRU  \\
                                             & (dispersed)         \\
\hline
\end{tabular}
\footnotetext[0]{\hspace{-\parindent}\emph{PRU} primary repeat
unit, \emph{SRU} secondary repeat unit, \emph{TRU} tertiary repeat
unit}
\end{table}

\subsection{Alphoid Higher Order Repeat Units in Human
and Chimpanzee Y Chromosome}

\subsubsection{Riddled HOR Scheme with 45 Distinct Alphoid Monomers
in Human Y Chromosome}

\begin{figure*}[t]
\includegraphics[width=\textwidth]{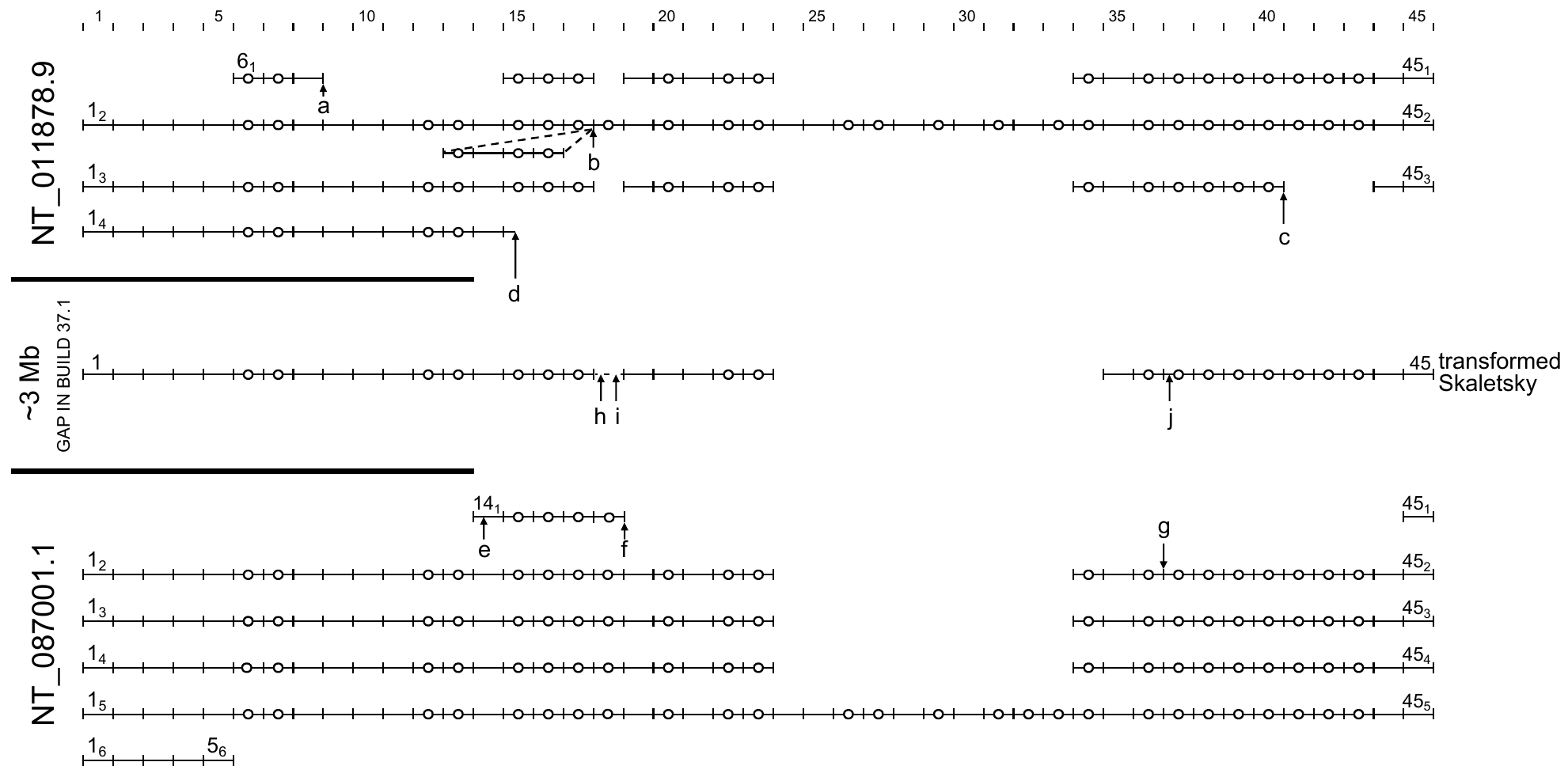}
\caption{\label{f1} \textsf{\small{Schematic presentation of aligned monomer
structure of 45mer alphoid HOR (consensus length 7662 bp) in human chromosome
Y (Build 37.1). This method of schematic presentation of HOR sequences is
self--evident if one compares Fig.\ \ref{f1} and Supplementary Table 1.
\emph{Top} enumeration of columns corresponding to 45 constituent consensus
monomers (enumerated Nos. 1 to 45) in consensus HOR. (For simplicity, only
every f\mbox{}ifth number is shown.) Each HOR copy is presented by a
\emph{bar} in the corresponding column numerated at the \emph{top}. Monomers
from dif\mbox{}ferent HOR copies corresponding to the same monomer from
consensus HOR are presented by \emph{bars} in the same column corresponding
to its enumeration at the \emph{top}. For example, in the f\mbox{}irst HOR
copy the f\mbox{}irst monomer corresponds to monomer No. 6 in consensus HOR
and is presented by a bar at position of 6th column (denoted by $6_1$), the
second monomer in the f\mbox{}irst HOR copy corresponds to monomer No. 7 in
consensus HOR and is presented by a \emph{bar} at the position of 7th
column\dots, the fourth monomer in the f\mbox{}irst HOR copy corresponds to
monomer No. 15 in consensus HOR and is presented by a \emph{bar} at the
position of 15th column\dots, and the last monomer in the f\mbox{}irst HOR
copy (the 23rd) corresponds to the monomer No. 45 in consensus HOR and is
presented by a \emph{bar} at the position of 45th column. \emph{Upper panel}:
HOR copies in contig NT\_011878.9. \emph{Lower panel}: HOR copies in contig
NT\_087001.1. \emph{Middle panel}: The 5941 bp secondary periodicity sequence
from \citet{skaletsky} mapped into alphoid monomers $\{m\}$. For mapping of
$\{w\}$--monomers from \citet{skaletsky} into $\{m\}$--monomers, see the text
and Supplementary Tables 2–-4. Open circle: pJ$\alpha$ motif (essential part)
in alpha monomers. The $m05$ monomer from the last incomplete HOR copy
($5_6$) in contig NT\_087001.1 is followed by alpha satellite monomeric
region (not shown here). \emph{a} After $m08$: 210 bp insertion (no
similarity to HOR monomers); \emph{b} after $m13$--$m16$ duplication
(inserted after $m17$) there are two insertions: 170 bp insertion
(dif\mbox{}fering in 19 bases from $m24$ and $m34$ as the closest monomers
from HOR) and 168 bp insertion (dif\mbox{}fering in 20 bases from $m28$ as
the closest monomer from HOR); \emph{c} after $m40$: 278 bp insertion (no
similarity to HOR monomers); \emph{d} after the f\mbox{}irst 34 bases from
$m15$: end of the contig NT\_011878.9; \emph{e} the last 166 bases of $m14$:
start of the contig NT\_087001.1; \emph{f} after $m17$: 311 bp insertion (no
similarity to HOR monomers); \emph{g} after $m36$: 171 bp insertion
(dif\mbox{}fering in 13 bases from $m23$ as the closest monomer from HOR);
\emph{h}, \emph{i} two deletions in $w20$; \emph{j} 53 bp nonalphoid
insertion in $w29$}}}
\end{figure*}

The largest repeat array in human Y chromosome assemblies studied
here is the major alphoid HOR array and, as will be shown here,
strongly diverges from the chimpanzee alphoid HOR. For this
reason, we f\mbox{}irst present our results for alphoid HORs. In
the contig NT\_087001.1 in centromere of human chromosome Y and in
NT\_011878.9 in the pericentromeric region on the proximal side of
p arm (DYZ3 locus), we identify the peripheral segments of the
major block of alphoid HOR array. In the spacing between these two
contigs lies a large central section of this HOR array. This
spacing of $\sim$3 Mb was not sequenced so far in the Build 37.1
assembly. The GRM results for alphoid monomer structure of the two
peripheral HOR segments are shown in Fig.\ \ref{f1} and
Supplementary Table 1. In Fig.\ \ref{f1}, we use a method of
schematic presentation described by \citet{rosandic06}.

In each of these two segments we identify 45 distinct alphoid
monomers, denoted $m01,\dots ,m45$, arranged head--to--tail in the
same orientation and mutually diverging by $\sim$20\%. The
consensus length of this 45mer HOR is 7662 bp. Here, an alphoid
monomer is assigned as constituent of HOR if it appears in at
least two HOR copies at a very low mutual divergence. Consensus
sequences of monomers forming HOR are shown in Supplementary Table
2. In both the contigs, the consensus sequences of monomers
constituting HOR are equal, ref\mbox{}lecting the fact that they
are two peripheral segments of the same HOR array (Table
\ref{t4}).
\begin{table}[h]
\centering \caption{\label{t4}\textsf{\small{Riddled pattern with
variety of number of monomers in human alphoid HOR copies (Build
37.1 assembly).}}}
\begin{tabular}{lp{0.1cm}lp{0.1cm}l}
\hline \multirow{3}{*}{HOR copy no.} & & \multicolumn{2}{l}{No. of monomers} \\
\cline{3-5}           & & Counting distinct  & & Counting all  \\
                      & & monomers           & & monomers      \\
\hline
1\footnotemark[1]   & & 23    & & 23 \\
2                   & & 45    & & 51 \\
3                   & & 31    & & 31 \\
4\footnotemark[1]   & & 14    & & 14 \\
5\footnotemark[1]   & & 6     & & 6  \\
6                   & & 35    & & 36 \\
7                   & & 35    & & 35 \\
8                   & & 35    & & 35 \\
9                   & & 45    & & 45 \\
10\footnotemark[1]  & & 5     & & 5  \\
\hline
\end{tabular}
\footnotetext[1]{Truncated at the start or end of the contig.
Copies No.\ 1-–4 are from contig NT\_011878.9. Copies No.\ 5--10
are from contig NT\_087001.1}
\end{table}

Divergence between monomers in individual HOR copies and the
corresponding consensus monomers is very low (on the average
0.3\%). However, the HOR structure is characterized by some
pronounced monomer deletions and insertions, giving a riddled
pattern (Table \ref{t4}) due to a variety of lengths of HOR copies
(Fig.\ \ref{f1}). We f\mbox{}ind monomer deletions in seven HOR
copies, monomer insertions in two, and nonalphoid insertions of
0.2 to 0.3 kb in three HOR copies. (In some HOR copies there are
multiple insertions and/or deletions.)

Two out of ten HOR copies contain the 10--alphoid--monomer
subsequence $m24, \dots ,m33$ (Fig.\ \ref{f1}). These ten monomers
are positioned between the monomers $m23$ and $m34$. Distance
between the two highly identical 10--alphoid--monomer subsequences
is $\sim$3 Mb.

The other 35 alphoid monomers from 45 distinct alphoid monomers in the
peripheral region of major alphoid HOR form a subsequence, consisting of two
segments, additionally riddled at some positions. Each of these 35 alphoid
monomers appears in three or more HOR copies (Fig.\ \ref{f1}). If we delete
the 10--alphoid--monomer subsequence from the 45mer, we obtain a 5957 bp
35mer, which is similar to the secondary periodicity sequence of 5941 bp
reported in \citet{skaletsky}.

Discussing relationship of the initially reported 5.7 and 6.0 kb
repeat units, Tyler--Smith and Brown proposed that one HOR unit is
derived from the other, although more complex explanations, with
both units derived from a third unknown HOR unit were considered
as possible \citep{tyler87}. It was considered as very unlikely
that the 6.0 kb unit arose from a 5.7 kb unit by addition of two
alphoid monomers, because results excluded the possibility that
the two additional alphoid monomers in the 6.0 kb unit are
duplications of any monomers contained in the 5.7 kb unit
\citep{tyler87}. Therefore, the favored hypothesis was that the
shorter, 5.7 kb HOR unit arose from the longer 6.0 kb HOR unit by
deletion of two alpha monomers. Extending similar considerations
to the present case, the 35mer in internal centromere region could
be considered as arising from 45mer by deletion of ten alphoid
monomers which are all distinct from the monomers in 35mer. This
is consistent with a general view \citep{warburton96a} that a type
of polymorphism found in alphoid arrays can be related to HOR
units that dif\mbox{}fer by an integral number of alphoid
monomers.

Divergence pattern provides an additional evidence that ten
additional alphoid monomers $m24,\dots ,m33$ are constituents of
major HOR. Mutual divergence between these ten monomers is similar
to their mean divergence with respect to the other 35 monomers
(Table \ref{t5}).
\begin{table}[h]
\centering \caption{\label{t5}\textsf{\small{Average divergence
between two subsets of alphoid monomers from 45mer HOR copies.}}}
\begin{tabular}{lp{0.2cm}l}
\hline Monomer comparison & & Divergence (\%) \\
\hline
10 vs. 10 & & $\sim$19 \\
10 vs. 35 & & $\sim$20 \\
35 vs. 35 & & $\sim$21 \\
\hline
\end{tabular}
\footnotetext[0]{\hspace{-\parindent}10 denotes the subset of ten
new monomers $m24,\dots m33$}
\footnotetext[0]{\hspace{-\parindent}35 denotes the subset of 35
monomers $m01,\dots m23$ and $m34,\dots m45$}
\end{table}

\subsubsection{Suprachromosomal Family Assignment of Monomers
in 45mer HOR}

Studies of sequence comparison of alpha satellite monomers in
human chromosomes revealed 12 types of monomers, forming
f\mbox{}ive suprachromosomal families (SFs), which descend from
two basic subsets of monomers, A and B: to the subset A belong the
SF types J1, D2, W4, W5, M1, and R1, and to the subset B belong
J2, D1, W1, W2, W3, and R2
\citep{romanova,warburton96a,alexandrov}. We determine the SF
assignments of monomers constituting alphoid HOR by pairwise
comparison between every monomer from HOR to every of 12 SF
consensus monomers from \citet{romanova}. A $45 \times 12$
divergence matrix is constructed between 45 monomers from HOR and
12 SF consensus monomers from \citet{romanova}. To each monomer
from HOR we assign the SF classif\mbox{}ication of the most
similar SF consensus monomer. In this way we f\mbox{}ind that, out
of forty--f\mbox{}ive monomers from HOR, forty monomers are of M1
type (in most cases the second lowest divergence corresponds to
R2, and in three cases the M1 and R2 divergences are equal), and
f\mbox{}ive are of R2 type (in these cases the second lowest
divergence corresponds to M1 type).

The dif\mbox{}ferences between A and B subsets are, in general,
concentrated in a small region which matches functional protein
binding sites for pJ$\alpha$ in subset A and for CENP--B in subset
B \citep{romanova}. Analyses of human genome have indicated that a
CENP--B box appears in the subset B monomers (in about 60\% of
B--type monomers) and is absent in the subset A monomers; while
the pJ$\alpha$ motif would occur only in some of monomers from the
subset A and not in the subset B monomers \citep{romanova}.

After determining the SF classif\mbox{}ication of monomers in consensus HOR,
we investigate the appearance of CENP--B box and pJ$\alpha$ motif in these
monomers. We f\mbox{}ind that the pJ$\alpha$ motif (essential part) is
present in 55\% of ten new alphoid monomers and similarly, in 57\% of the
other 35 monomers, while the CENP--B box is completely absent (Fig.\
\ref{f1}). Consensus HOR has a robust pJ$\alpha$ distribution, containing 25
pJ$\alpha$ motif copies. All alphoid monomers in consensus HOR are
signif\mbox{}icantly more similar to pJ$\alpha$ motif than to the CENP--B
box: the mean deviation is 0.6 bp for the pJ$\alpha$ motif and 4.7 bp for the
CENP--B box, ref\mbox{}lecting that the absence of pJ$\alpha$ motif in some
of monomers from 45mer HOR can be attributedmostly to a single nucleotide
mutation within an initially pJ$\alpha$ motif.

Since the pJ$\alpha$ motif is essential for protein binding, an
interesting question is whether the monomers with and without
pJ$\alpha$ motif have dif\mbox{}ferent sequence divergences. In
this respect, pairwise divergence among 45 monomers shows no
dependence on the presence or absence of the pJ$\alpha$ motif.

It should be noted that HOR copies in chromosome Y are the only
reported case where pJ$\alpha$ motif is present and CENP--B box
absent.

In this connection, we note a unique case of 13mer HOR (2214 bp
consensus length) in chromosome 5, which contains neither CENP--B
box nor pJ$\alpha$ motif \citep{rosandic06}.

\subsubsection{Alignment of Peripheral and Internal Human HOR Copies}

Let us now compare our consensus HOR for the peripheral parts of
major HOR alphoid block (DYZ3 locus) (Supplementary Table 2) to
the 5941 bp secondary periodicity sequence in its internal part
reported by \citet{skaletsky} which corresponds to the sequence
gap between the contigs NT\_011878.9 and NT\_087001.1 in the Build
37.1 assembly.

F\mbox{}irst, we fragment the 5941 bp sequence from
\citet{skaletsky} into 35 constituent alpha monomers, denoted
$w01,\dots ,w35$ (Supplementary Table 3). We f\mbox{}ind a
peculiar feature of this secondary periodicity sequence: two of
its constituent monomers, $w20$ and $w29$, exhibit sizeable length
deviation from the alpha satellite consensus length of 171 bp: the
alphoid monomer $w20$ has a length of 104 bp (i.e., 67 nucleotides
are deleted with respect to consensus alpha monomer length) while
the monomer $w29$ is 224 bp long, containing a 53 bp nonalphoid
insertion with respect to consensus alpha monomer.

To align the internal monomer sequence $\{w\}$ (Supplementary
Table 3) to the peripheral monomer sequence $\{m\}$ (Supplementary
Table 2), we shift the start position of alpha monomers $m01,
m02,\dots ,m45$, obtaining the sequence denoted by $n01, n02,\dots
,n45$ (Table \ref{t6}). The 35 alphoid monomers from the sequence
$\{w\}$ are aligned to 35 out of 45 monomers $\{n\}$ (Table
\ref{t6} and Supplementary Table 4). The sequences $n26,\dots
,n35$ have no counterpart in the $\{w\}$ sequence which
corresponds to internal part of major alphoid HOR from
\citet{skaletsky}.\begin{table}[!hb] \centering
\caption{\label{t6}\textsf{\small{Transformation between monomer
sets $\{m\}$ and $\{n\}$ and alignment between alphoid monomer
sets  $\{w\}$ and  $\{n\}$}}}
\begin{tabular}{l}
\hline
\emph{Transformation} \\
$n01(169) = m44(\dots 113) + m45(056\dots)$ \\
$n02(166) = m45(\dots 110) + m01(056\dots)$ \\
\dots \\
$n45(170) = m43(\dots 114) + m44(056\dots)$ \\
\emph{Alignment} \\
$w01 = n01$ \\
\dots \\
$w25 = n25$ \\
$w26 = n36$ \\
\dots \\
$w35 = n45$ \\
\hline
\end{tabular}
\footnotetext[0]{\hspace{-\parindent}For def\mbox{}inition of
monomers $\{n\}$ and $\{w\}$ see Supplementary Tables 3 and 4. In
the transformation from $\{m\}$ to $\{n\}$ the notation $m44(\dots
113)$ denotes the last 113 bases in $m44$, $m45(056\dots)$ denotes
the f\mbox{}irst 56 bases in $m45$, and so on (Supplementary Table
4). In alignment between $\{n\}$ and $\{w\}$ the 35 alphoid
monomers from the sequence $\{w\}$ are aligned to 35 out of 45
monomers from the sequence $\{n\}$. Here, the only
signif\mbox{}icant dif\mbox{}ferences appear between $w20$ and
$n20$ (due to the presence of deletion in $w20$), and between
$w29$ and $n39$ (due to presence of insertion in $w29$). The
monomers $n26,\dots n35$ have no counterpart in the set $\{w\}$
which corresponds to the internal part of major alphoid HOR}
\end{table}

\subsubsection{Global Repeat Map for Riddled Alphoid HOR
and Characteristic HOR--Signature in Human Chromosome Y}

\begin{figure}[!h]
\includegraphics[width=\columnwidth]{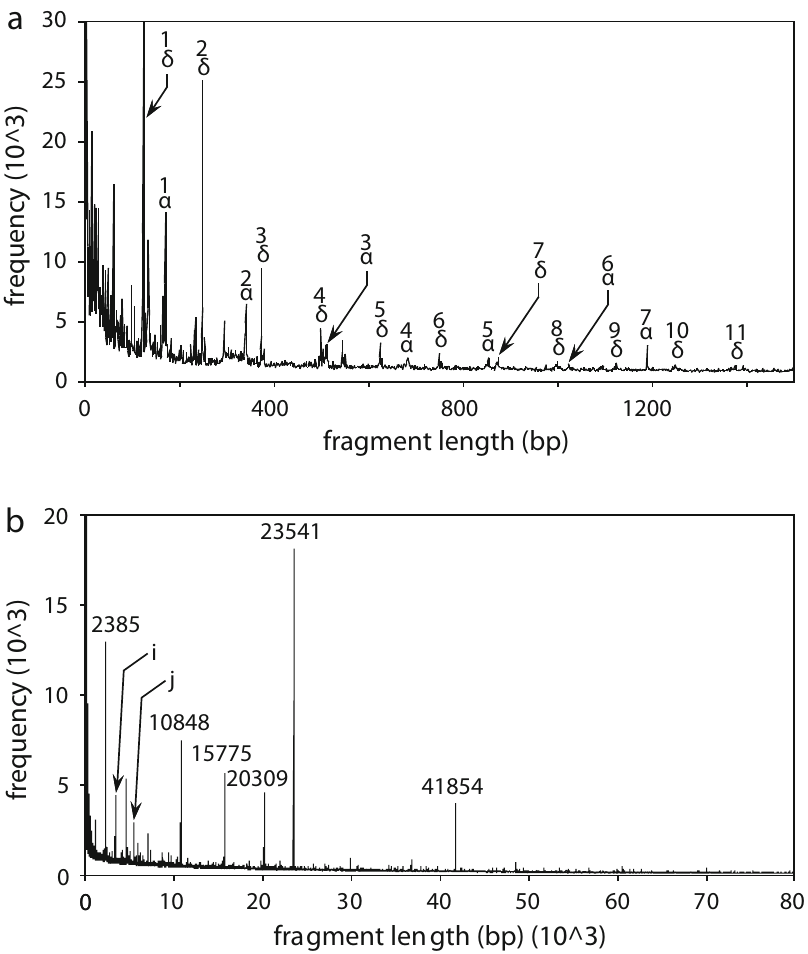}
\caption{\label{f2} \textsf{\small{GRM diagram for Build 37.1
genomic assembly of human chromosome Y for the intervals of
fragment lengths: \textbf{a} 0–-1500 bp. There are two pronounced
tandem arrays with repeat units below 1.5 kb: the alphoid tandem
repeat with alpha satellite repeat unit of 171 bp and the
overlapping tandem repeat with repeat unit of 125 bp. The peaks at
multiples of alphoid monomer repeat unit 171 bp, $n \cdot$171 bp,
are denoted by $n\alpha$. The peaks at multiples of 125 bp repeat
unit, $n\cdot$125 bp, are denoted by $n\delta$. \textbf{b}
0–-80000 bp. Pronounced peaks above 2 kb are denoted by the
corresponding fragment lengths. The most pronounced peaks are
approximately at 2385, 10848, 15775, 20309, 23541, and 41584 bp.
\emph{Arrow i}: peak corresponding to 715mer. \emph{Arrow j}: peak
corresponding to 1123mer. For description of peaks see the text}}}
\end{figure}

To investigate more closely the major alphoid HOR array in human
chromosome Y, we compute the GRM diagram for genomic sequence of Y
chromosome (Fig.\ \ref{f2}). The most pronounced peaks in this
diagram correspond to following tandem repeats in chromosome Y:
the alphoid repeats (GRM peaks at multiples of the $\sim$171 bp
repeat unit), the 125 bp repeats (GRM peaks at multiples of the
125 bp repeat unit), GRM peaks at multiples of 5 bp repeat unit
and GRM peaks corresponding to $\sim$20.3 kb repeat unit. In
addition, there are nine pronounced GRM peaks at repeat lengths
above 2000 bp.

\begin{figure}[h!]
\includegraphics[width=\columnwidth]{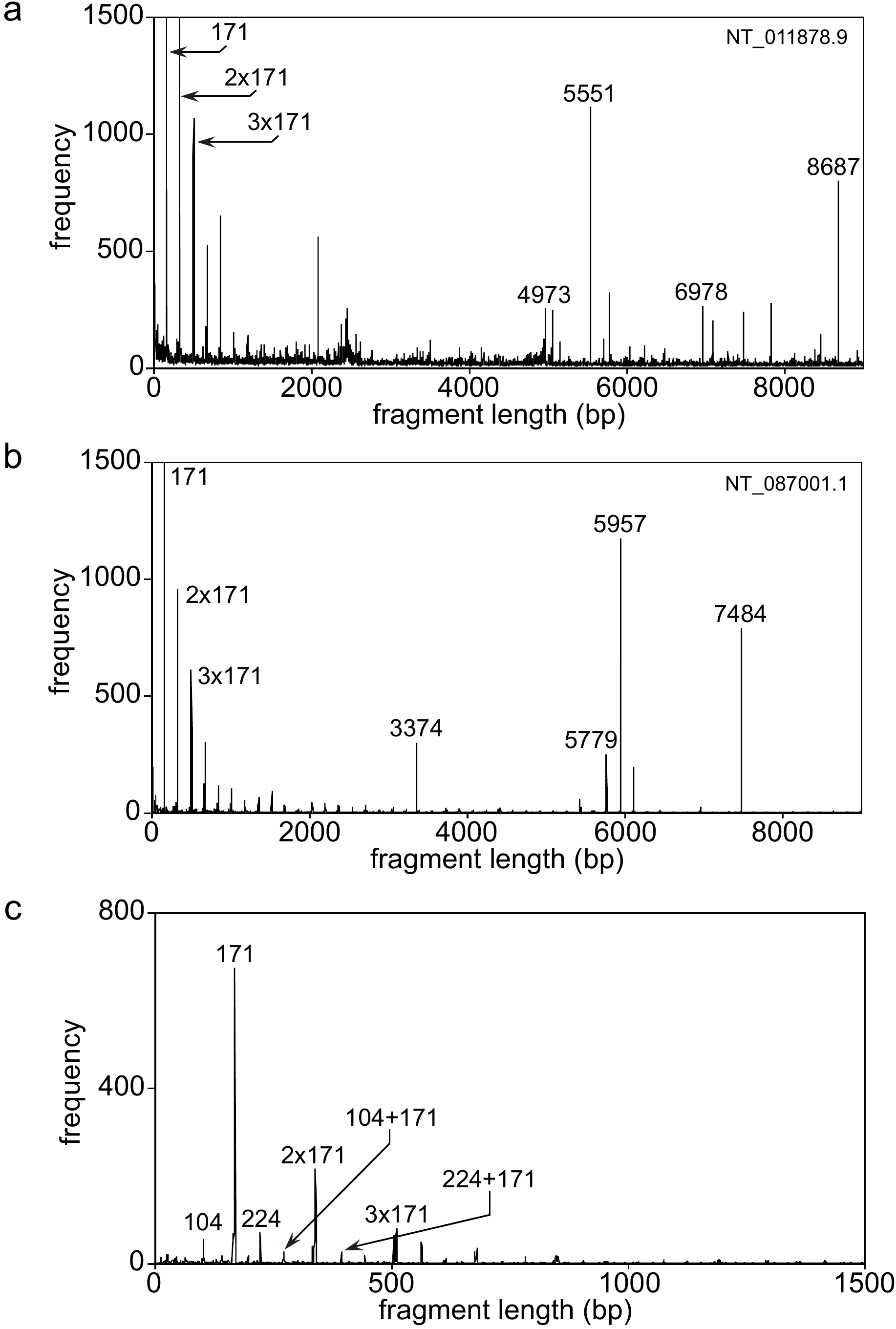}
\caption{\label{f3} \textsf{\small{GRM diagrams for sequences in
contigs containing alphoid HOR in chromosome Y: \textbf{a}
NT\_011878.9, \textbf{b} NT\_087001.1, and \textbf{c} secondary
periodicity sequence for internal part of major interior alphoid
HOR block (genomic sequence from \citet{skaletsky})}}}
\end{figure}
Here, we perform detailed study for alphoid HOR repeat sequence.
Analyzing partial contributions to GRM diagram of chromosome Y
from individual contigs we f\mbox{}ind that the largest frequency
contributions to alphoid HOR peaks are arising from the contigs
NT\_011878.9 and NT\_087001.1. The relevant intervals of fragment
lengths for these two contigs are shown in Fig.\ \ref{f3}a and b,
respectively. In both the f\mbox{}igures peaks at approximate
multiples of basic repeat length $\sim$171 bp are decreasing with
increasing multiple orders. That is a natural trend for tandem
repeats. However, we do not f\mbox{}ind a peak corresponding to
the HOR length, which for regular HORs in other chromosomes
appears at their consensus lengths. This is because the Build 37.1
assembly of chromosome Y encompasses only peripheral tails of
major HOR array and those exhibit sizeable riddling in both
relevant contigs, as shown in the monomer structure of peripheral
HOR copies in Fig.\ \ref{f1}. For these riddled HOR copies there
is no dominating consensus length and therefore no peak
corresponding to consensus length is present. Instead, the GRM
diagram shows more intricate HOR--related peaks which characterize
riddled alphoid HOR copies. These peaks will be referred to as GRM
HOR--signature. Most pronounced GRM HOR--signature peaks of
riddled HOR pattern in peripheral regions of major alphoid HOR in
chromosome Y are at the lengths shown in Fig.\ \ref{f3}a, b. These
characteristic fragment lengths are fully consistent with the
riddled HOR structure from Fig.\ \ref{f1}.

As an example, let us consider the largest GRM HOR signature peak at 5551 bp,
characterizing HOR pattern in NT\_011878.9. This peak arises from approximate
repeat of the $1_3$--$14_3$ subsequence at the position of the $1_4$--$14_4$
subsequence. The distance $l$ between the corresponding bases in these two
subsequences (Table \ref{t7}) is equal to a distance between monomers $1_3$
and $1_4$ (Fig.\ \ref{f1} and Supplementary Table 1).
\begin{table}[h]
\centering \caption{\label{t7}\textsf{\small{Contributions to the
fragment length 5551 bp alphoid GRM HOR--signature peak for human
Y chromosome}}}
\begin{tabular}{lp{0.2cm}l}
\hline
Length (bp) & & Distance      \\
\hline
2896        & & $1_3 –- 17_3$    \\
848         & & $19_3 –- 23_3$   \\
1194        & & $34_3 –- 40_3$   \\
278         & & Nonalphoid insertion  \\
335         & & $44_3 –- 45_3$   \\
$\sum 5551$ & &               \\
\hline
\end{tabular}
\end{table}

\begin{figure*}[t]
\includegraphics[width=0.9\textwidth]{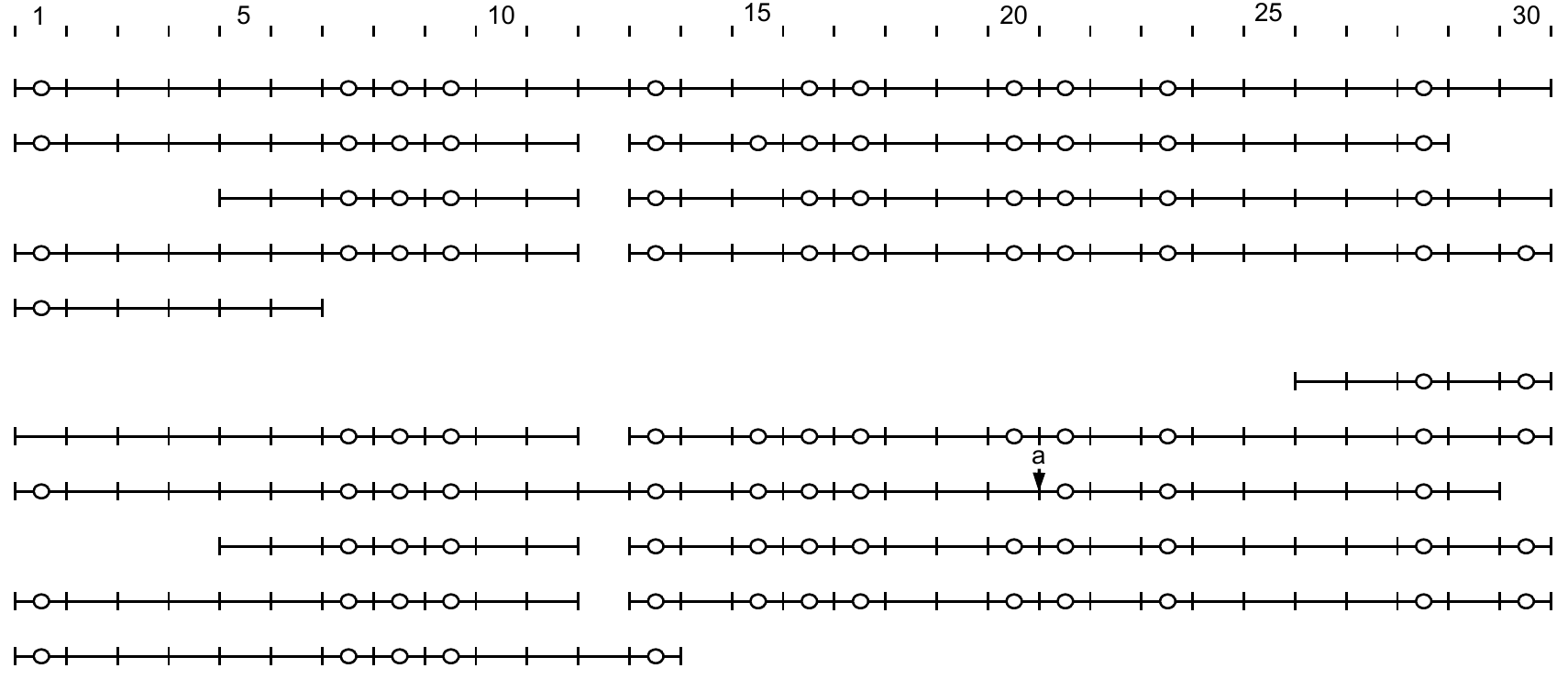}
\caption{\label{f4} \textsf{\small{Schematic presentation of
aligned monomer structure of 30mer alphoid HOR (consensus length
5066 bp) in chimpanzee chromosome Y (Build 2.1, contig
NW\_001252921.1). \emph{Top row} enumeration of 30 constituent
alpha monomers from consensus HOR. \emph{Upper panel}: HOR copies
in interval 264–-20019. \emph{Lower panel} reverse complement of
HOR copies in interval from 20618–-42459. After monomer No.\ 20
(label \emph{a}): 41 bp insertion (no similarity to monomers in
30mer). For comparison with human alphoid HOR see Fig.\ \ref{f1}.
\emph{Open circle} pJ$\alpha$ motif (essential part) in alpha
monomers}}}
\end{figure*}

Therefore, the GRM diagram shows a pronounced peak at the 5551 bp
fragment length, ref\mbox{}lecting the riddling structure of HORs.
Similarly, we interpret all the other HOR--signature peaks which
characterize riddling in HOR copies from Fig.\ \ref{f1}.

In addition to GRM computation for Build 37.1 sequence of
chromosome Y, let us comment on the GRM HOR--signature related
irregularity (monomers $w20$ and $w29$) in the interior region of
major alphoid HOR array in chromosome Y (Supplementary Tables 3,
4). Figure \ref{f3}c displays GRM diagram computed for the 5941 bp
secondary periodicity sequence from \citet{skaletsky}. Here again,
we see the main pattern of monomer multiples $\sim$171, $\sim 2
\times 171$, $\sim 3 \times 171$ bp,\dots with decreasing
frequencies for increasing multiples. In addition, we obtain two
weak subsequences of peaks, at fragment lengths $\sim$104 bp,
$\sim$ ($104 + 171$ bp), $\sim$ ($104 + 2 \times 171$ bp), \dots
and at $\sim$224 bp, $\sim$($224 + 171$ bp), $\sim$($224 + 2
\times 171$ bp),\dots These two additional weak subsequences are
due to two distorted monomers in the 35mer periodicity (HOR)
sequence that we deduced from the HOR genomic sequence in
\citet{skaletsky}: the alphoid monomer $w20$ has a length of 104
bp (i.e., 67 nucleotides are deleted with respect to consensus
monomer) while the monomer $w29$ has the length 224 bp, containing
a 53 bp nonalphoid insertion with respect to consensus monomer.
Such deletions/insertions in two distant alphoid monomers within
HOR are absent in the peripheral regions of major HOR array in
chromosome Y, i.e., they are absent in Build 37.1 assembly.
Therefore, GRM diagrams of these regions (Fig.\ \ref{f3}a, b) do
not have these two additional weak subsequences of peaks. This
actualizes the interest for future extension of Build assembly to
the region of sequence gap of $\sim$3 Mb between the contigs
NT\_011878.9 and NT\_087001.1.

\subsubsection{Riddled 30mer HOR Scheme in Chimpanzee
Chromosome Y}

Applying GRM to the chimpanzee chromosome Y, we f\mbox{}ind two
30mer HOR arrays in chimpanzee contig NW\_001252921.1 (NCBI Build
2.1), positioned one after another (with a gap of 599 bp in
between) at the front part of the contig. The f\mbox{}irst HOR,
truncated at the start of the contig is referred to as direct. In
fact, it seems to be a truncated tail of a major HOR block
positioned in unsequenced domain in front of the contig
NW\_001252921.1. We f\mbox{}ind that the reverse complement of the
second HOR array is highly identical to the f\mbox{}irst HOR
array, and therefore this second HOR array is referred to as
reverse complement. This indicates that the direct and reverse
complement HOR arrays are positioned on the opposite arms of a
palindrome.

Our results for detailed monomer scheme of these two peripheral
HOR arrays, which are reverse complement to each other, are shown
in Fig.\ \ref{f4} and Supplementary Table 5. The consensus length
of 30mer HOR unit is 5066 bp (consensus sequence in Supplementary
Table 6).

\begin{figure}[t]
\includegraphics[width=\columnwidth]{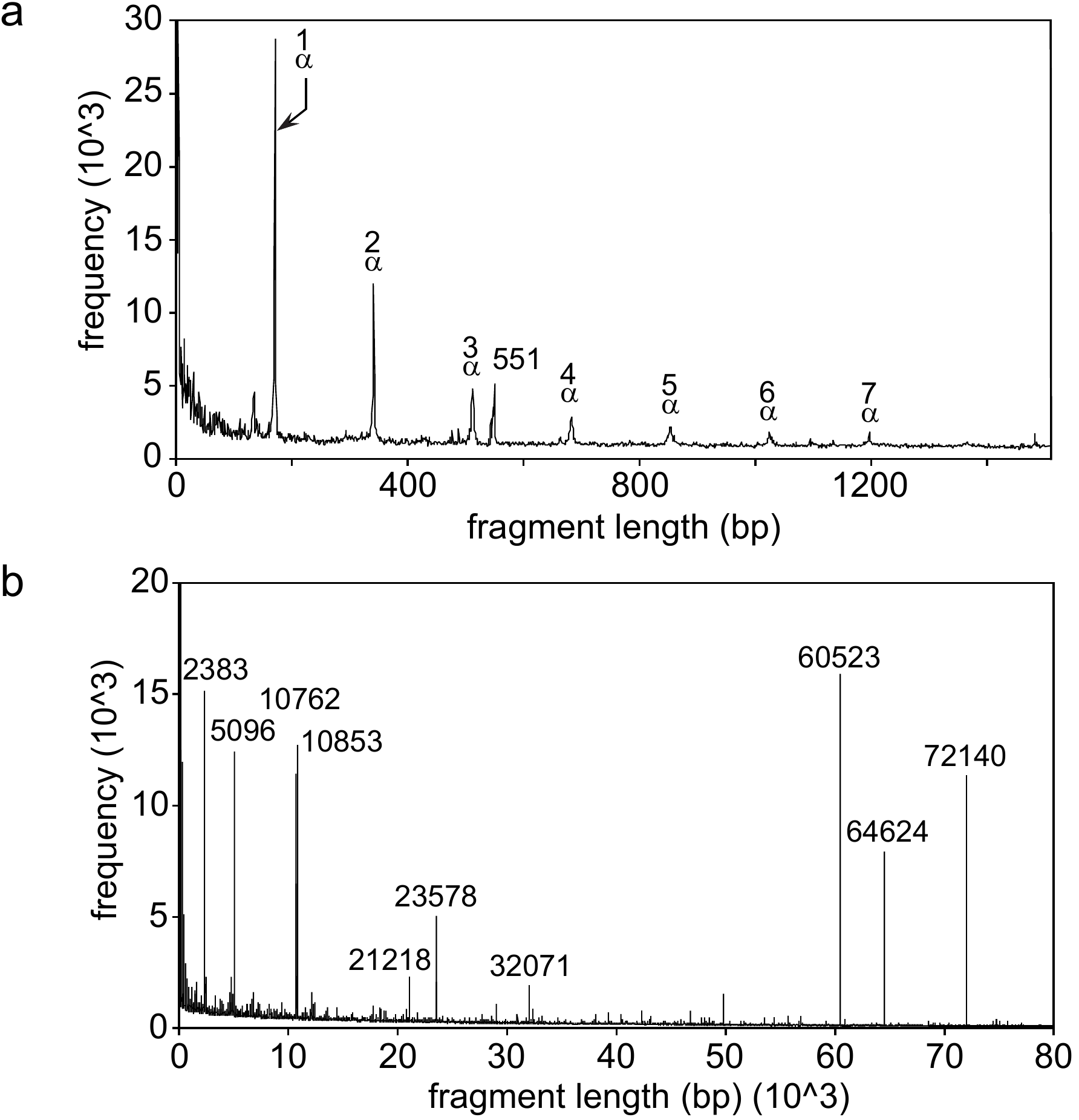}
\caption{\label{f5} \textsf{\small{GRM diagram for Build 2.1
genomic assembly of chimpanzee chromosome Y for intervals of
fragment lengths: \textbf{a} 0–-1500 bp. There is only one
pronounced tandem array with repeat units in the interval between
0.1 and 1.5 kb: the alphoid tandem repeat with alpha satellite
repeat unit of 171 bp. The peaks at multiples of alphoid monomer
repeat unit 171 bp, $n\cdot$171 bp, are denoted by $n\alpha$.
\textbf{b} 0–-80000 bp. Pronounced peaks above 2 kb are denoted by
the corresponding fragment lengths. The most pronounced peaks
above 1.5 kb are approximately at 2383, 5096, 10762, 10853, 21218,
23578, 32071, 60523, 64624, and 72140 bp. For description of peaks
see the text}}}
\end{figure}
\begin{figure}[b]
\includegraphics[width=\columnwidth]{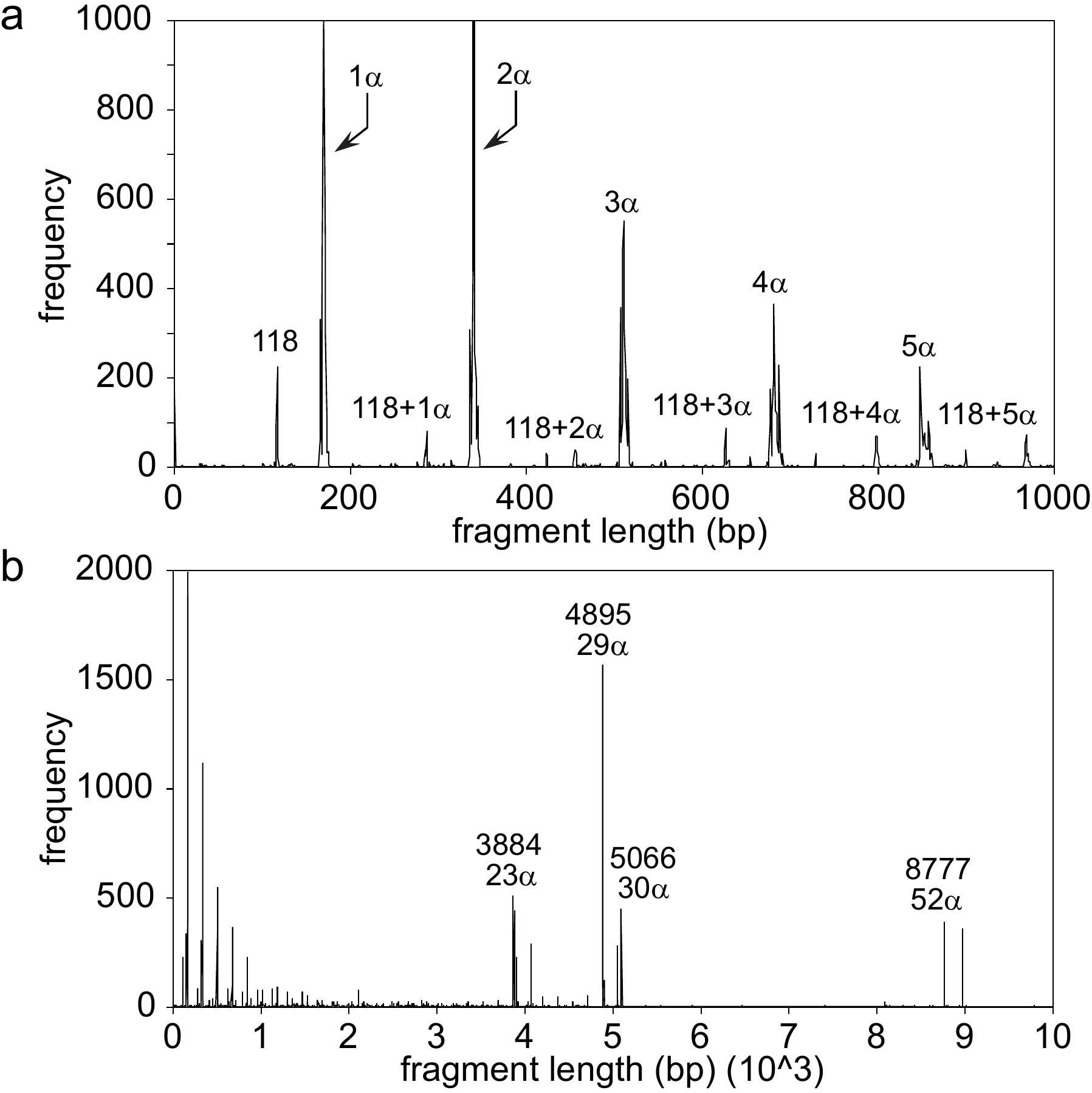}
\caption{\label{f6} \textsf{\small{GRM diagram for HOR containing
section from positions 1–-20019 bp in the chimpanzee contig
NW\_001252921. Intervals of fragment lengths: \textbf{a} 0–-1000
bp, \textbf{b} 0–-10000 bp. For description of peaks see the
text}}}
\end{figure}

In GRM diagram of the whole chimpanzee Y chromosome (Fig.\
\ref{f5}), the peak at 5066 bp fragment length is much weaker than
the near--lying 5096 bp peak of another repeat structure (see
Tables \ref{t2}, \ref{t3}) and is therefore overshadowed. For this
reason, we compute the GRM diagram selectively for alphoid
HOR--containing section of genomic sequence at the start of contig
NW\_001252921.1 (positions 1--20019) (Fig.\ \ref{f6}). In Fig.\
\ref{f6}, in the length interval between 0.1 and 1 kb there are
pronounced peaks approximately at multiples of alphoid monomer
repeat unit 171 bp (Fig.\ \ref{f6}a), in analogy to Fig.\
\ref{f5}a for the whole chimpanzee chromosome Y. Furthermore, the
HOR--signature peaks are clearly seen in Fig.\ \ref{f6}b as
pronounced peaks at 5066 bp ($\sim 30 \times 171$ bp, denoted as
$30\alpha$), 4895 bp ($\sim 29 \times 171$ bp, denoted as
$29\alpha$), 3884 bp ($\sim 23 \times 171$ bp, denoted as
$23\alpha$), and 8777 bp ($\sim 52 \times 171$ bp, denoted as
$52\alpha$). These HOR--signature peaks can be also deduced
directly from HOR structure from Fig.\ \ref{f4} and Supplementary
Table 5.

For example, the 8777 bp ($52\alpha$) HOR--signature peak arises
from the approximate repeat of the $1_2$--$4_2$ subsequence at
position of the $1_4$--$4_4$ subsequence (the $1_3$--$4_3$
subsequence is missing due to riddling) (Table \ref{t8}). Distance
between the corresponding bases in these two subsequences is equal
to the distance between monomers $1_2$ and $1_4$.
\begin{table}[h]
\centering \caption{\label{t8}\textsf{\small{Contributions to
fragment length 8777 bp alphoid GRM HOR--signature peak for
chimpanzee Y chromosome}}}
\begin{tabular}{lp{0.2cm}l}
\hline
Length (bp) & & Distance      \\
\hline
1868        & & $1_2 –- 11_2$    \\
2683        & & $13_2 -– 28_2$   \\
1198        & & $5_3 –- 11_3$    \\
3028        & & $13_3 –- 30_3$   \\
$\sum 8777$ & &               \\
\hline
\end{tabular}
\end{table}

Similarly, we interpret all the other pronounced HORsignature
peaks in Fig.\ \ref{f6}b. The frequencies of these peaks are
sizably smaller than of peaks arising from some other tandem
repeats and therefore are overshadowed in Fig.\ \ref{f5} for the
whole chimpanzee Y chromosome. We note that the HOR--signature
peaks at 3884, 4895, 5066, and 8777 bp are the only
signif\mbox{}icant GRM peaks above 1.5 kb in Fig.\ \ref{f6}b.

Some peaks from GRM diagram for the whole chromosome Y (Fig.\
\ref{f5}) are missing in GRM diagram for the HOR section in Fig.\
\ref{f6}a. For example, the peak at 551 bp from Fig.\ \ref{f5}a is
missing in Fig.\ \ref{f6}a, because the repeat unit of 551 bp is
positioned outside of the HOR--section of genomic sequence
included in Fig.\ \ref{f6}a.

In addition to the equidistant multiple alphoid peaks, in the GRM
diagram in Fig.\ \ref{f6}a there is a family of weaker equidistant
peaks at fragment length 118, $118 + \alpha$, $118 + 2\alpha$,
$118 + 3\alpha$,\dots (like in Fig.\ \ref{f5}, here $\alpha$,
2$\alpha$, 3$\alpha$,\dots denote multiples of alpha monomer
length $\sim$171 bp). This weak equidistant family of repeat
lengths is based on the 118 bp peak. The origin of this peak is
that one of monomers within HOR, $m25$, is truncated, with size
reduced from the standard value $\sim$171 to 118 bp. (Observe that
we f\mbox{}ind an analog appearance of additional bands based on
monomers of irregular length, 104 and 224 bp, for two human
monomers in 35mer alphoid HOR in the interior part of HOR array.)

\subsubsection{Comparison of Alpha Satellite Monomers in Human 45mer
and Chimpanzee 30mer HORs}

Computing divergence between 45 human consensus alpha monomers
from consensus 45mer HOR and 30 chimpanzee consensus alpha
monomers from consensus 30mer HOR (Supplementary Table 7) we see
that due to scattering of divergences and the absence of any small
divergence, none of chimpanzee monomers can be assigned to a
particular human monomer (Supplementary Table 8). In the whole
human--chimpanzee divergence matrix the lowest divergence value is
12\%, appearing in a few cases only (Table \ref{t9}).
\begin{table}[h]
\centering \caption{\label{t9}\textsf{\small{Illustration of
divergences of human monomers $m01$ and $m24$ with respect to 30
chimpanzee monomers}}}
\begin{tabular}{lp{0.2cm}lp{0.2cm}l}
\hline \multirow{2}{*}{Human monomer} & & No.\ of
chimpanzee & & \multirow{2}{*}{Divergence (\%)} \\
    & & monomers & & \\
\hline
$m01$   & & Two   & & 21 \\
$m01$   & & Four  & & 22 \\
$m01$   & & Three & & 23 \\
$m24$   & & Three & & 12 \\
$m24$   & & Three & & 13 \\
$m24$   & & One   & & 14 \\
\hline
\end{tabular}
\footnotetext[0]{\hspace{-\parindent}This means, for example, that
the lowest divergences between $m01$ (human) monomer and each of
30 chimpanzee monomers is 21\% (with respect to two chimpanzee
monomers), 22\% (with respect to four chimpanzee monomers), etc.}
\end{table}
The mean value of the lowest human--chimpanzee divergence for each
human monomer is 17\% (Supplementary Table 8). The absence of
identity between particular human and chimpanzee monomers from
alphoid HORs is also seen from the mean values of divergences in
Table \ref{t10}.
\begin{table}[h]
\centering \caption{\label{t10}\textsf{\small{Comparison of mean
values of human and chimpanzee consensus monomer divergences}}}
\begin{tabular}{lp{0.2cm}l}
\hline                    & & Divergence (\%) \\
\hline
45 Human vs.\ 45 human              & & 19    \\
30 Chimpanzee vs.\ 30 chimpanzee    & & 21    \\
45 Human vs.\ 30 chimpanzee         & & 23    \\
\hline
\end{tabular}
\footnotetext[0]{\hspace{-\parindent}45 human denotes the set of
consensus alpha monomers from human 45mer HOR, and 30 chimp from
chimpanzee 30mer HOR}
\end{table}

On the other hand, we f\mbox{}ind that alpha monomers in 30mer
HORs in chimpanzee Y chromosome are predominantly of M1 SF type,
similarly as alpha monomers in 35mer/45mer HORs in human
chromosome Y. Accordingly, similarly as for human Y chromosome,
monomers in chimpanzee Y chromosome are also characterized by the
presence of pJ$\alpha$ motif and the absence of CENP--B box (Fig.\
\ref{f4}). As already noted, the human Y chromosome was the only
known case where pJ$\alpha$ motif is present and CENP--B box
absent and now we see that the chimpanzee Y chromosome shares this
feature.

As to the degree of riddling, the human HOR is more riddled than
the chimpanzee HOR. In particular, the human HOR has more
insertions than the chimpanzee HOR, which is ref\mbox{}lected in
their respective GRM HOR signature.

\subsubsection{Peculiarities of Alphoid HOR in Human Y Chromosome}

We show that HOR structure in the peripheral regions of the major
alphoid block in human chromosome Y is more complex than the
previously reported structure for the internal region. In this
computational study, we identify and fully characterize the
peripheral region, in particular f\mbox{}inding ten new monomers
constituting alphoid HOR copies, dif\mbox{}ferent from the known
35 constituentmonomers, giving evidence for the presence of 45mer
in the peripheral region of HOR array. Furthermore, while 33 out
of 35 constituting alphoid monomers in HOR copies in the interior
HOR region are highly homologous to the corresponding monomers in
the peripheral region, we f\mbox{}ind that the remaining two
monomers in the interior region have a sizeable deletion and
nonalphoid insertion, respectively, with respect to the
corresponding monomers from the peripheral region. The study of
these riddled HOR copies may be valuable for understanding
possible sources of genomic diversity, but also has the potential
to provide useful markers for medical, population, and forensic
genetic studies, and may give a route for identifying mechanisms
of DNA sequence evolution.

Some peculiarities studied in this work regarding the major
alphoid HOR that may shed some new light at the mysteries of human
Y chromosome are:

The 33 consensus monomers from the peripheral HOR structure are
highly identical to the aligned 33 monomers of previously reported
secondary periodicity sequence from \citet{skaletsky}. On the
other hand, we f\mbox{}ind peculiar dif\mbox{}ferences: the 10mer
alphoid sequence, inserted in the peripheral HOR structure, is
absent in the reported internal structure; and in the previously
reported internal secondary periodicity structure one constituent
alphoid monomer has a sizeable deletion (67 bp) and the other a
sizeable nonalphoid insertion (53 bp) accompanied by clustered
substitutions of 11 bases with respect to the peripheral HOR
structure.

The highly identical alphoid 10mer insert appears in both
peripheral regions of major HOR, but was not reported so far in
the internal centromere region between the two peripheral regions.

The peripheral regions of major HOR alphoid block reveal
coexistence: on one hand, very low divergence between the aligned
constituent alpha monomers from dif\mbox{}ferent HOR copies
(average divergence$\sim$0.3\%) and, on the other hand, pronounced
riddling due to deletions and insertions of alpha monomers and/or
due to insertions of nonalphoid segments. The HOR copies in
chromosome Y are the only known case where the pJ$\alpha$ motif is
present and CENP--B box absent.

The major alphoid HOR in Y chromosome exhibits more deletions and
insertions of alphoid monomers and highly distorted insertions
than HORs in other chromosomes.

\subsubsection{Dif\mbox{}ference Between Humans and Chimpanzees Alphoid
HOR Repeat Units}

The number of dif\mbox{}ferent monomers constituting HOR in human
Y chromosome (45 monomers in the peripheral sections of major HOR
array, and 35 monomers in the interior section) is
dif\mbox{}ferent than in the chimpanzee genome (30 monomers).

HOR pattern in the sequenced domain in Build 37.1 assembly
(peripheral region) is characterized by substantial riddling,
which is more pronounced in human than in chimpanzee genome.

All alpha satellite monomers constituting major human 35/45mer HOR
are dif\mbox{}ferent from monomers constituting chimpanzee 30mer
HOR by$\sim$20\%, which is comparable to divergence between
monomers within a single HOR copy.

The lengths of major alphoid HOR arrays in human and chimpanzee
are widely dif\mbox{}ferent, $\sim$3 and $\sim$1 Mb, respectively.

\subsection{Other Human and Chimpanzee Tandem, HOR
and Regularly Dispersed Repeat Arrays Based on Large Repeat Units}

Besides the alphoid HOR, in human Build 37.1 and chimpanzee Build
2.1 Y chromosome assemblies we f\mbox{}ind over 20 other large
repeat units (Tables \ref{t1}, \ref{t2}, \ref{t3}). Some of large
repeat units appear both in human and in chimpanzee genomic
assembly, and some in human only or in chimpanzee only. We
describe here some pronounced repeats identif\mbox{}ied from GRM
diagrams (labeled a in Tables \ref{t1}, \ref{t2}). The remaining
repeats (denoted b in Tables \ref{t1}, \ref{t2}) are described in
Supplementary information.

\subsubsection{Chimpanzee $\sim$550 bp Primary Repeat Unit, $\sim$1652 bp
3mer HOR Secondary Repeat Unit, and $\sim$23578 bp Tertiary Repeat
Unit}

In the GRM diagram for chimpanzee Y chromosome in the length
interval between 100 and 1500 bp, besides the major peaks
associated with alphoid HOR and tandem repeat based on the 125 bp
repeat unit, there is additional pronounced peak at $\sim$550 bp
(Fig.\ \ref{f5}a). Using GRM, we f\mbox{}ind that this peak arises
due to the appearance of 3mer HOR copies constituted from
three$\sim$550 bp monomers, denoted $mc01$ $mc02$ and $mc03$.
These monomers are mutually diverging by $\sim$8\%, while
dif\mbox{}ferent 3mer HOR copies mutually diverge by only
$\sim$1\%. About eight times smaller divergence between 3mer
copies then between individual monomers within each 3mer are a
signature of HOR. However, these HOR copies are not in tandem, in
contrast to previously known HOR structures; instead, they are
dispersed with rather regular spacings. Consensus sequences of
three monomers $mc01$ $mc02$ and $mc03$, determined from
NW\_001252921.1 (using key string AGGTACTG) are given in
Supplementary Table 9. The main contributions to the $\sim$550 bp
GRM peak arise from the array of $\sim$550 bp monomers within each
3mer copy.

\begin{table}[h]
\centering \caption{\label{t11}\textsf{\small{Dispersed 3mer HOR
copies based on $\sim$550 bp monomer in chimpanzee Y chromosome}}}
\begin{tabular}{llll}
\hline \multirow{2}{*}{Contig} & HOR copy &
\multirow{2}{*}{Direction} & Monomers \\
& start position &  & in HOR copy \\
\hline
NW\_001252921.1     & 618296    & RC    & $mc03$ $mc02$ $mc01$  \\
                    & 1021470   & D     & $mc01$ $mc03$         \\
                    & 1044498   & D     & $mc01$ $mc02$ $mc03$  \\
                    & 1330195   & RC    & $mc03$ $mc02$ $mc01$  \\
                    & 1353792   & RC    & $mc03$ $mc02$ $mc01$  \\
                    & 1757803   & D     & $mc01$ $mc02$ $mc03$  \\
                    & 1781391   & D     & $mc01$ $mc02$ $mc03$  \\
                    & 2063781   & RC    & $mc03$ $mc02$ $mc01$  \\
                    & 2087364   & RC    & $mc03$ $mc01$         \\
                    & 2490023   & D     & $mc01$ $mc03$         \\
                    & 2513051   & D     & $mc01$ $mc02$ $mc03$  \\
                    & 2798724   & RC    & $mc03$ $mc01$         \\
NW\_001252926.1     & 232825    & D     & $mc01$ $mc02$ $mc03$  \\
                    & 516623    & RC    & $mc03$ $mc02$ $mc01$  \\
                    & 540165    & RC    & $mc03$ $mc02$ $mc01$  \\
NW\_001252919.1     & 328953    & D     & $mc01$ $mc02$ $mc03$  \\
                    & 574371    & RC    & $mc03$ $mc02$ $mc01$  \\
NW\_001252925.1     & 922846    & D     & $mc01$ $mc02$ $mc03$  \\
                    & 1197882   & RC    & $mc03$ $mc02$ $mc01$  \\
NW\_001252915.1     & 955834    & RC    & $mc03$ $mc02$ $mc01$  \\
\hline
\end{tabular}
\footnotetext[0]{\hspace{-\parindent}\emph{RC} denotes a HOR copy
having reverse complement sequence with respect to HOR copy
def\mbox{}ined as direct (D). In reverse complement HOR copy each
monomer is reverse complement with respect to direct monomer
sequence}
\end{table}
Performing the GRM analysis we f\mbox{}ind 20 dispersed HOR copies
(Table \ref{t11}). In addition, in four HOR copies in
NW\_001252921.1 one of three $\sim$550 bp monomers is deleted. In
NW\_001252921.1, we f\mbox{}ind dispersed highly identical 3mer
HORs, direct and reverse complement. HOR copies after the
f\mbox{}irst one are grouped into f\mbox{}ive pairs of 3mers:
\begin{center}
\begin{tabular}{lll}
D & S & D \\
R & S & R \\
D & S & D \\
R & S & R \\
D & S & D \\
\end{tabular}\end{center}
where D is the direct 3mer copy, R is the reverse complement 3mer
copy, and S is the spacing of $\sim$24 kb (see Table \ref{t11}).
(Three of 3mer copies in these pairs of 3mer copies are truncated
from three to two monomers.) Since the two 3mer copies in each
pair are separated by spacing S, there is no GRM peak at
$\sim$1.65 kb. Instead, this gives rise to a tertiary repeat unit,
with a $\sim$24 kb peak (more precisely $\sim$23578 bp) in the GRM
diagram.

We f\mbox{}ind even an approximate next higher pattern, three
copies of quartic repeat unit:
$$ \mbox{R S}_2\!\mbox{ D S D S}_1\!\mbox{ R S R S}_2\!\mbox{ D S D S}_1\!\mbox{ R S R S}_2\!\mbox{ D S D S}_1\!\mbox{ R} $$
where S$_2$ is spacing of $\sim$0.40 Mb, and S$_1$ spacing of $\sim$0.28 Mb
(see Table \ref{t11}). The length of this unit is $\sim$0.73 Mb. In
NW\_001252921.1, we f\mbox{}ind an array of three such quartic repeat units.
This would give rise to a GRM peak at $\sim$0.74 Mb fragment length
(computation is performed here up to 100 kb fragment lengths).

We note that in NW\_001252926.1 we f\mbox{}ind a $\mbox{D
S}_1\mbox{ R S R}$ subsection of the above pattern.

\subsubsection{Human $\sim$545 bp Primary Repeat Unit, $\sim$1641 bp 3mer
HOR Secondary Repeat Unit, and $\sim$23541 bp Tertiary Repeat
Unit}

The GRM peak at 545 bp is due to the $\sim$545 bp monomers,
organized in dispersed 3mer HOR copies of $\sim$1641 bp (Table
\ref{t12}). The distance between start positions of two 3mer
copies is again $\sim$24 kb, similar as in the chimpanzee Y
chromosome, giving rise to the appearance of $\sim$23541 bp peak
in GRM diagram.
\begin{table}[h]
\centering \caption{\label{t12}\textsf{\small{Dispersed 3mer HOR
copies based on $\sim$545 bp monomer in human Y chromosome}}}
\begin{tabular}{lp{0.1cm}lp{0.1cm}lp{0.1cm}l}
\hline \multirow{2}{*}{Contig} & & HOR copy & &
\multirow{2}{*}{Direction} & & Monomers \\
& & start position & &  & & in HOR copy \\
\hline
NT\_011903.12   & & 76992     & & RC    & & $m03$ $m02$ $m01$ \\
                & & 100533    & & RC    & & $m03$ $m02$ $m01$ \\
                & & 365459    & & RC    & & $m03$ $m02$ $m01$ \\
                & & 609306    & & D     & & $m01$ $m02$ $m03$ \\
NT\_011875.12   & & 9862260   & & D     & & $m01$ $m02$ $m03$ \\
                & & 9885800   & & D     & & $m01$ $m02$ $m03$ \\
                & & 9909341   & & D     & & $m01$ $m02$ $m03$ \\
NT\_086998.1    & & 185824    & & D     & & $m01$ $m03$       \\
\hline
\end{tabular}
\footnotetext[0]{\hspace{-\parindent}\emph{RC} denotes a HOR copy
having reverse complement sequence with respect to HOR copy
def\mbox{}ined as direct (D). In reverse complement HOR copy each
monomer is reverse complement with respect to direct monomer
sequence}
\end{table}

The 23541 bp repeat unit corresponds to previously reported 23.6
kb repeat units containing RMBY genes, but previously it was not
related to the 545 bp PRU \citep{skaletsky,warburton08}.

As seen, the human HOR pattern of sequenced Y chromosome contains
fewer copies than chimpanzees and is less symmetrically organized.
The human $\sim$545 bp monomers (denoted $m01$ $m02$ $m03$) are
similar to the chimpanzee $\sim$550 bp monomers (denoted $mc01$
$mc02$ $mc03$): divergence between the human 3mer HORs $m01$,
$m02$, and $m03$ and the chimpanzee 3mer HORs is $\sim$4\%, while
the divergence between of\mbox{}f--diagonal monomers (i.e., $m01$
vs. $mc02$, $m01$ vs. $mc03$,\dots) is $\sim$8\%. Only a small
subsection of $\sim$24 kb encompassing each human HOR copy is
similar to the corresponding section encompassing each chimpanzee
HOR copy (divergence less than 10\%), while the remaining part of
large spacings, of total length $\sim$2 Mb, strongly diverges
between human and chimpanzee. This gives a substantial
contribution to the overall human--chimpanzee divergence.
Furthermore, the subsequences of $\sim$24 kb human sequence are
scattered in various parts of chimpanzee Y chromosome.

\subsubsection{Human $\sim$2385 bp Primary Repeat Unit and $\sim$7155 bp
3mer HOR Secondary Repeat Unit}

The DAZ gene family, located in the AZFc region of Y chromosome,
is organized into two clusters and contains a variable number of
copies \citep{seboun,glaser,saxena00,fernandes}. A $\sim$2.4 kb
repeat unit in DAZ genes was reported by
\citep{skaletsky,warburton08}. Accordingly, the GRM peak at 2385
bp (Fig.\ \ref{f2}b) is due to tandem repeats with $\sim$2.4 bp
PRU in DAZ genes. Human DAZ repetitions are located in contig
NT\_011903.12 (positions 1346649 to 1361029, 1425263 to 1473290,
2977988 to 2997102, and 3050498 to 3086580), i.e., from position
25.3 to 27 Mb within the human Y chromosome.

Using GRM we classify the assembly of $\sim$2.4 kb monomers into
f\mbox{}ive monomer families (consensus sequences in Supplementary
Table 10). The average divergence between monomers of the same
family is below 1\%, while the average divergence between monomers
from dif\mbox{}ferent families is $\sim$11\%. The monomer family
with highest frequency of appearance has consensus length 2385 bp,
which determines the length of the 2385 bp GRM peak. This monomer
family forms a highly homologous monomeric tandem repeat, which is
present in DAZ2 and DAZ4 genes.

We f\mbox{}ind that the GRM peak at 7155 bp corresponds to 3mer
HOR composed of three variants of $\sim$2.4 kb DAZ repeat
monomers, denoted $m01$, $m02$, and $m03$ (the f\mbox{}irst three
consensus sequences from Supplementary Table 10). Computing the
GRM diagram of any of the 7155 bp copies we obtain two pronounced
peaks, at $\sim$2.4 and $\sim$4.8 kb, revealing the 3mer
character. We f\mbox{}ind that these 3mer HOR copies are present
in all four DAZ1--DAZ4 genes. Human DAZ genes contain 12 DAZ HOR
copies organized into four tandem arrays (DAZ1--DAZ4).

The $\sim$4757 bp peak in GRM diagram corresponds to the 2mer HOR
copies arising from 3mer HOR by deletion of one monomer from the
7155 bp secondary 3mer HOR unit. In GRM diagram of the 4757 bp
repeat copies, we obtain only one pronounced GRM peak, at
$\sim$2.4 kb, showing the 2mer character of 4757 bp repeat copies.
We f\mbox{}ind that such 2mer HOR copies are present in all four
DAZ1--DAZ4 genes.

\subsubsection{Chimpanzee $\sim$2383 bp Primary Repeat Unit and Absence
of Tandem of Higher Order Repeats}

The GRM peak at $\sim$2383 bp is due to tandem repeats with
$\sim$2.4 bp repeat unit in DAZ genes in chimpanzee Y chromosome.
Chimpanzee DAZ repetitions are located in contigs NW\_001252917.1
(positions 1109191 to 1130961 and 1259092 to 1280862) and
NW\_001252922.1 (positions 997017 to 1028356 and 1070171 to
1099128) that is at chromosome positions from $\sim$3.2 to 3.4 Mb
and from $\sim$11.2 to 11.3 Mb. Positions of the corresponding
subsequences widely dif\mbox{}fer in human and chimpanzee
chromosomes. Divergence between human and chimpanzee consensus
sequences is $\sim$5\%.

We f\mbox{}ind that the chimpanzee Y chromosome contains 3mer and
2mer HOR copies, similar to those for human Y chromosome, but with
one pronounced distinction: chimpanzee DAZ genes contain four DAZ
HOR copies, which are, unlike the case of human Y chromosome, not
organized into tandem but into dispersed HOR copies. Therefore,
there are no GRM peaks corresponding to HORs.

The presence of tandem of DAZ HOR copies in human and absence of
such tandem in chimpanzee Y chromosome provides an interesting
evolutionary distinction between human and chimpanzee Y
chromosomes.

\subsubsection{Human $\sim$3579 bp 715mer HOR Unit and 5 bp Primary
Repeat Unit}

The GRM peak at $\sim$3579 bp is due to a tandem of 28 repeat
copies in NT\_025975.2. These copies dif\mbox{}fer in lengths from
3544 to 3589 bp. The length 3579 bp has the highest frequency and
is equal to consensus length. Other copy lengths appear due to
deletion or insertion of 5 bp subsequences. Average divergence of
copies with respect to consensus sequence is $\sim$1\%. Due to
dif\mbox{}ferences in lengths of copies, the GRM peak at
$\sim$3579 bp is broadened (Fig.\ \ref{f2}b).

In the next step, we f\mbox{}ind a strong peak at the fragment
length 5 bp in GRM diagram for the 3579 bp consensus sequence. A
dominant key string for segmentation of the 3579 bp consensus
sequence into 5 bp fragments is ATTCC, which is the consensus
sequence of 5 bp primary repeat copies. Thus the 3579 bp repeat
unit is a 715mer HOR based on ATTCC primary consensus repeat unit.
Here 34\% of primary repeat 5 bp copies are equal to consensus,
38\% dif\mbox{}fer from consensus by one base, 21\% by two, 6\% by
three and 1\% by four bases.

This 3579 bp HOR corresponds to the previously reported 3584 bp
HOR \citep{skaletsky}.

\subsubsection{Absence of Chimpanzee HOR Unit Corresponding
to Human 3579 bp 715mer HOR Unit}

In the Build 2.1 assembly for chimpanzee Y chromosome we
f\mbox{}ind no analog of the human 3579 bp 715mer HOR unit.

\subsubsection{Human $\sim$5607 bp 1123mer HOR Unit and 5 bp Primary
Repeat Unit}

The 5607 bp peak corresponds to a new HOR, with 5607 bp SRU (5 bp
GGAAT PRU). The main contribution to this peak is from contig
NT\_113819.1. We identify a tandem of 11 copies, from position
496682 to 553881 (Supplementary Table 11) and determine the 5607
bp consensus sequence (Supplementary Table 12).

To investigate the structure of 5607 bp repeat unit, we compute
the GRM diagram of its consensus sequence. Using 8 bp key string
ensemble, we obtain the GRM diagram characterized by a set of GRM
peaks at fragment lengths of 5 bp and its multiples (Supplementary
Fig.\ \ref{f1}a), revealing the underlying 5 bp PRU. However, the
reciprocal distribution of GRM peaks shows deviation from the
exponential distribution expected due to random mutations of
fragments of multiple orders at KSA recognition sites. This
deviation is due to the fact that the length of key strings in the
ensemble is larger than the repeat unit. This is shown by
computing the GRM diagram by using the 3 bp key string ensemble,
shorter than the 5 bp PRU (Supplementary Fig.\ \ref{f1}b). In that
case the reciprocal distribution of GRM peaks corresponding to the
5607 bp consensus sequence indeed follows exponential
distribution, as expected.

The 5607 bp HOR consensus unit consists of 1123 pentamer copies.
Out of these copies, 353 are identical to GGAAT which is the
primary repeat consensus. The mean divergence between 5 bp
consensus GGAAT and pentamer copies that are not identical to
consensus is $\sim$30\%. Dif\mbox{}ferences are mostly due to
substitutions. There are only a few indels: two copies have
1--base insertion, one has 2--base insertion, ten have 1--base
deletion and one has 2--base deletion.

\subsubsection{Absence of Chimpanzee HOR Unit Corresponding
to the Human 5607 bp HOR Unit}

In the Build 2.1 assembly for chimpanzee Y chromosome we
f\mbox{}ind no repeat unit corresponding to human 5607 bp HOR
unit.

\subsubsection{Chimpanzee 10853 bp Primary Repeat Unit and 64624 bp
Secondary Repeat Unit}

The GRM peak at 10853 bp is due to a tandem in NW\_001252917.1
(eight copies), with repeat unit consensus length 10853 bp. The
10853 bp consensus sequence is given in Supplementary Table 15.
The third copy in this tandem is distorted: truncated after the
f\mbox{}irst 6399 bases and followed by a large insertion, so that
the total length of truncated third copy and neighboring insertion
amount to the combined length of 21218 bp. The structure of the
eighth copy is distorted similarly as the third copy, leading
again to a $\sim$21 kb combined length.

Distance between the corresponding bases in neighboring copies
(except those involving the third copy) is $\sim$10853 bp, giving
rise to the 10853 bp GRM peak.

Distance between the start of the 6399 bp subsection of the third
copy and the start of the fourth copy is 21218 bp, giving rise to
the 21218 bp GRM peak. Distance from the end of the second copy
(which has no counterpart in the truncated third copy) to the end
of the fourth copy is $10853 + 21218$ bp $= 32071$ bp, giving rise
to the 32071 bp GRM peak.

The copies No. 1, 2, and 4--7 are identical up to 1\%, while the
copies No. 3 and 8 have similar truncation and additional
insertion. Therefore, the copies No. 1--5 form a secondary repeat
HOR copy of the approximate length $2 \times 10853 + 21218 + 2
\times 10853$ (precise value 64624 bp). The last three copies in
tandem, No. 6--8, represent the f\mbox{}irst three copies
belonging to the second 64624 bp HOR copy.

The insertion after the truncated third copy in chimpanzee tandem
repeat with 10853 PRU 21218--6399 bp $=$ 14819 bp is also present
in the human Y chromosome as a tandem of two repeat units
(divergence $\sim$ 4\%) in contig NT\_011903.12. Because these
repetitive units are mutually reverse complement, GRM diagram for
human chromosome Y does not show this peak.

\subsection{Summary of Human--Chimpanzee Divergence Due
to Repeats Based on Large Repeat Units}

We determine approximately the number of bases which are
dif\mbox{}ferent in repeat arrays of human and chimpanzee Y
chromosome using a simple formula:
\begin{equation}
d = \sum_i d_i  =\sum_i\Big( \min
(l_{i,\mathrm{hum}},l_{i,\mathrm{chimp}}) \cdot p_i + l_i\Big).
\end{equation}

Here, $l_{i,\mathrm{hum}}$ and $l_{i,\mathrm{chimp}}$ are sums of
lengths over all copies of the $i$th's human and chimpanzee repeat
unit, respectively; $\min
(l_{i,\mathrm{hum}},l_{i,\mathrm{chimp}})$ is the smaller of two
lengths $l_{i,\mathrm{hum}}$ and $l_{i,\mathrm{chimp}}$; $l_i =
|l_{i,\mathrm{hum}} - l_{i,\mathrm{chimp}}|$; and $p_i$ is
divergence between human and chimpanzee repeat unit \emph{i}. In
this way, we include contributions to human--chimpanzee divergence
both from substitutions and indels.

\begin{figure}[!h]
\includegraphics[width=\columnwidth]{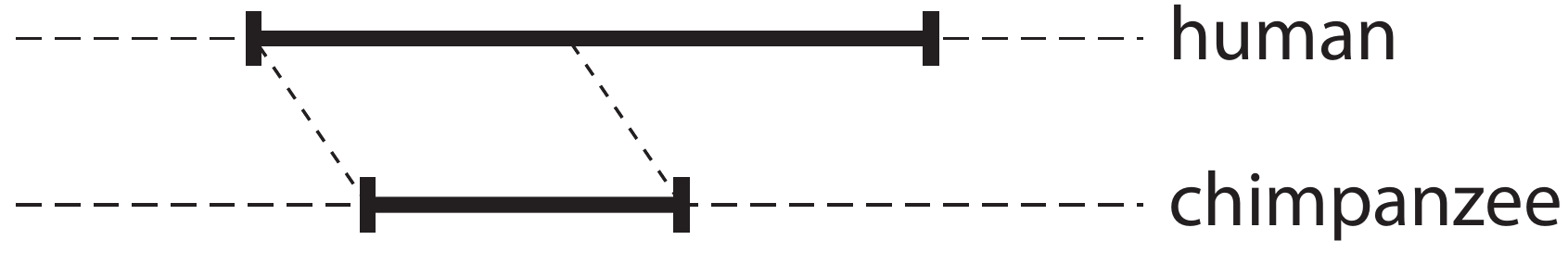}
\put(-200,40){$\sim l_{i,\mathrm{chimp}}$} \put(-130,40){$l_i$}
\put(-178,13){$l_{i,\mathrm{chimp}}$}\put(-210,18){$p_i$}\put(-170,55){$l_{i,\mathrm{human}}$}
\caption{\label{f7} \textsf{\small{Schematic presentation of
applying the formula for calculation of human–-chimpanzee
divergence for the case of a large repeat unit (major alphoid
HOR)}}}
\end{figure}

For example, in the case of alphoid HOR in Y chromosome (repeat
No. 1 from Tables \ref{t1}, \ref{t2}, \ref{t3}) we have:
$l_{1,\mathrm{hum}} = 3048138$ bp, $l_{1,\mathrm{chimp}} =
1042459$ bp, $l_1 = 2005679$ bp, $p_1 = 0.20$, giving $d_1 =
2.214.171$ bp (Fig.\ \ref{f7}). With respect to the sequence of
larger alphoid HOR, of the length $l_{1,\mathrm{hum}}$, this
corresponds to an approximate divergence $100 \cdot d_1/
l_{1,\mathrm{hum}} = 72.6\%$.

Summing over all repeats ($i= 1,2,\dots $) from Tables \ref{t1},
and \ref{t3}, we obtain a summary number of dif\mbox{}ferent bases
between human and chimpanzee large repeats: $d \sim 3.4$ Mb
(3378539 bp). The corresponding divergence with respect to all
repeats from Tables \ref{t1}, \ref{t2}, and \ref{t3} is:
\begin{equation}
\mbox{div}(\mbox{rep}) = 100 \cdot \frac{d}{L},
\end{equation}
where the summary length of all repeats from Tables \ref{t1},
\ref{t2}, and \ref{t3} is $L = 4848892$ bp.

Thus, we obtain divergence with respect to repeat sequences
included in Tables \ref{t1}, \ref{t2}, and \ref{t3}:
\begin{equation}
\mbox{div}(\mbox{rep}) \approx 70\%.
\end{equation}
If we smear out divergence over the whole Build sequence of length
$L_{\mathrm{as}} = 25$ Mb, we obtain the overall divergence with
respect to assembly length:
\begin{eqnarray}
\mbox{div}(\mbox{Build}) &=& 100 \cdot \frac{d}{L_{\mathrm{as}}} \\
\mbox{div}(\mbox{Build}) & \approx & 14 \%.
\end{eqnarray}

This estimate of overall divergence due to repeats based on large
repeat units should be additionally increased due to overall
estimates of approximately 1--2\% divergence for nonrepeat
sequences.

Both the human and the chimpanzee Y chromosome sequences are still
incomplete; in human chromosome $\sim$25 Mb out of total length of
$\sim$59 Mb was sequenced. Thus, a greater contiguity at several
genomic regions is desired to reach more precise conclusions
regarding human--chimpanzee divergence. However, the main body of
results will probably stand, because, in general, nonsequenced
gaps are rich in repeat structures. It should be noted that a
whole--genome comparison of chimpanzee and human revealed an
increased divergence in the terminal 10 Mb of the corresponding
chromosomes, consistent with general association between increased
divergence rates and location near the chromosome ends
\citep{mikkelsen,pollard06a}. In general, and in accordance with
\citet{gibbs}, it can be expected that unsequenced regions of
repeat elements, that are dif\mbox{}f\mbox{}icult to align, might
for the whole Y chromosome somewhat increase the presently
estimated divergence of 14\% for the sequenced part.
Def\mbox{}initive studies of genome evolution will require
high--quality f\mbox{}inished sequences \citep{mikkelsen}.

An interesting question is how much the observed sizeable
divergence can be generalized to the whole genome. In this sense,
we have started a systematic study of human--chimpanzee divergence
due to large repeats in other chromosomes.

We see a tendency that large repeat units in humans are on average
larger and copy numbers greater than those in chimpanzees. This is
in accordance with previous observation that microsatellites in
humans are on average longer than those in chimpanzees
\citep{vowles}.

We identify large repeat units which contribute substantially to
divergence between humans and chimpanzees. Our results indicate
that alphoid HOR and most of characteristic tandem repeats with
large repeat units (some present only in human and not in
chimpanzee Y chromosome, or some vice versa) have been created
after the human--chimpanzee separation, while only a smaller
number of tandems with large repeat units (present both in human
and in chimpanzee Y chromosome at low mutual divergence) originate
from a common ancestor that predated the human--chimpanzee
separation. This is in accordance with previous observations in
some other chromosomes that alpha satellite subsets found in great
apes and humans are in general not located on their corresponding
homologous chromosomes \citep{jorgensen92,warburton96b}; for
example, the alpha satellite subset on human chromosome 5 is a
member of SF 1, while the homologous chimpanzee chromosome belongs
to SF 2 \citep{haaf97,haaf98}. It was pointed out that this
implies that the human--chimpanzee sequence divergence has not
arisen from a common ancestral repeat, but instead represents
initial amplif\mbox{}ication and homogenization of distinct
repeats on homologous chromosomes (nonorthologous evolution).

\citet{haaf97} discussed the propositions for homogenization of
alpha satellites. Homogenization processes appear to proceed in
localized, short--range fashion that leads to formation of large
domains of sequence identity \citep{tyler87,durfy89,warburton90}.
Genomic turnover mechanisms (molecular drive;
\citep{dover1,dover2}) must be at work that spread and homogenize
individual variant repeat units throughout arrays and throughout
populations \citep{haaf95}. However, the mechanisms by which this
concerted evolution occurs seem unclear, although several genomic
turnover mechanisms such as unequal crossing over between repeats
of sister chromatids \citep{smith}, sequence conversion
\citep{baltimore}, sequence transposition \citep{calos},
translocation exchange \citep{krystal}, and disproportionate
replication \citep{hourcade,spradling,lohe} have been observed to
be active in certain genomes.

Previous FISH studies support the conclusion that the localization
of SF 3 alpha satellite is substantially conserved, while alpha
satellite sequences belonging to families 1 and 2 are not shared
by the corresponding chimpanzee homologs
\citep{daiuto,archidiacono}. Here we f\mbox{}ind that, although
the SF 4 which is composed of M1 alpha satellite monomers
constituting human and chimpanzee alphoid HORs in Y chromosomes is
conserved, both the alpha satellite monomers in human and
chimpanzee HORs and the HOR lengths are widely dif\mbox{}ferent.

It was pointed out that it is not known whether evolutionary
important mutations predominantly occurred in regulatory sequences
or coding regions
\citep{king,mcconkey00,mcconkey02,olson,carroll}. Preliminary data
suggested that gene expression patterns of human brain might have
evolved rapidly \citep{enard,caceres,uddin,dorus}.

Comparative genomic analyzes strongly indicated that the marked phenotypic
dif\mbox{}ferences between humans and chimpanzees are likely due more to
changes in gene regulations then to modif\mbox{}ications of genes themselves
\citep{king,pollard06a,pollard06b,popesco,prabhakar}. The gene regulatory
evolution hypothesis proposes that the striking dif\mbox{}ferences between
humans and chimpanzees are due to gene expression: the change of pattern and
timing of turning genes on and of\mbox{}f.

\citet{pollard06b} identif\mbox{}ied $\sim$100 bp short genomic
regions that are highly conserved in vertebrates, but show
signif\mbox{}icantly accelerated substitution rates on human
lineage relative to chimpanzee \citep{pollard06a,pollard06b}. Many
of these Human Accelerated Regions (HARs), characterized by dense
clusters of nucleotide substitutions, are associated, in
particular, with the nervous system, reproductive system, and
immune system.

Detailed studies have indicated that forces other than selection
for random mutations that increase f\mbox{}itness in
specif\mbox{}ic functional elements may be at play in strongly
accelerated regions \citep{pollard06a}. There is a possibility
that changes in the accelerated regions result from a combination
of multiple evolutionary processes, perhaps including biased gene
conversion and a selection--based process \citep{pollard06a}.

Here, we f\mbox{}ind another type of accelerated regions: for some
repeat arrays we f\mbox{}ind dramatic evolutionary acceleration of
repeat pattern, from monomeric arrays in chimpanzee to HOR
organization of repeat arrays in human Y chromosome, i.e., the
rapid onset of unequal crossing over in human lineage. Such region
of accelerated evolution of HOR pattern will be referred to as
human accelerated HOR region (HAHOR).

The hallmark of evolutionary shift of function is sudden change in
a region of genome that previously has been conserved
\citep{pollard06b}. The function of sets of genomic regulatory
sequences has been previously compared to electronic
microprocessing: they process the information contained in a set
of regulatory elements into the corresponding pattern of gene
expression. It was noted that one of basic ways how the regulatory
genomic features are related to evolutionary processes is the
recruitment of existing regulatory pathways into newly evolving
context \citep{gierer98,tautz,pires}. These processes follow the
rules of nonlinear interactions. These, in turn, allow for sudden
or very fast changes resulting from the accumulation of rapidly
succeeding small steps with self--enhancing features. Furthermore,
mechanisms of bifurcation and de novo pattern formation may lead,
for instance, to strikingly dif\mbox{}ferent developments in parts
of an initially near--uniform area. Thus, in general, small causes
can result in big ef\mbox{}fects \citep{gierer04}. Finally we note
a possibility that accelerated large repeat units and HAHORs could
have a functional role of new categories of long--range regulatory
elements \citep{noonan}.

\section{Conclusion}

In this study, we identify and analyze tandem repeats, HORs and
regularly dispersed repeats in chimpanzee and human. For the
f\mbox{}irst time we report a dozen new large repeats in
chimpanzee and several new large repeats in human genome.
Comparing the corresponding repeats based on large repeat units in
human and chimpanzee we f\mbox{}ind substantial contribution to
the human--chimpanzee divergence from these repeats, approximately
70\% divergence with respect to repeat arrays based on large
repeat units. Smearing out these dif\mbox{}ferences in large
repeats over the whole sequenced assemblies, human Build 37.1 and
chimpanzee Build 2.1, i.e., by neglecting divergence between other
segments of genome sequences, we obtain an overall
human--chimpanzee divergence between sequenced assemblies of
approximately 14\%. This numerical estimate far exceeds the
available earlier numerical estimates for human--chimpanzee
divergence.

Our results are in accordance with recent publication by
\citet{hughes10} where it was shown by overall comparison that the
human and chimpanzee MSYs dif\mbox{}fer radically.

We explicitly identify, analyze, and compare a dozen of large
repeats which give a substantial contribution to human--chimpanzee
divergence.

We f\mbox{}ind in humans several HAHORs on human lineage relative
to chimpanzee, containing HOR structures, in particular the
alphoid HORs, the $\sim$2.4 kb DAZ repetitions and the $\sim$15.8
kb repetitions. On the other hand, in chimpanzee genome we
f\mbox{}ind a chimpanzee--accelerated HOR region (CAHOR) based on
$\sim$550 bp PRU.

While the HARs discovered previously
\citep{pollard06a,pollard06b,popesco,prabhakar,pollard09} were
HARs characterized by short dense clusters of nucleotide
substitutions, the HAHORs found in this work are characterized by
higher--order organization extended over larger genomic stretches.

Our results show explicitly that large repeat units and HORs
provide substantial contribution to the human--chimpanzee
divergence.

\section{GRM Analysis}

GRM analysis was performed using novel GRM code, which is
available upon request.

\begin{acknowledgments}
Authors are grateful to Martin Kreitman and Chris Tyler--Smith for
very helpful comments and suggestions.
\end{acknowledgments}

\end{document}